\documentclass[aps,prb,preprint,superscriptaddress,longbibliography]{revtex4-2}

\usepackage{amsmath,latexsym}
\usepackage{xcolor}
\usepackage{graphicx}
\usepackage{caption}
\usepackage{color}
\usepackage{dcolumn}
\usepackage{bm}
\usepackage{CJK}
\usepackage{braket}
\usepackage[font=small,labelfont=bf]{caption}
\usepackage{hyperref}

\makeatletter
\renewcommand*{\@fnsymbol}[1]{\ensuremath{\ifcase#1\or \dagger\or  \ddagger\or *\or
\mathsection\or \mathparagraph\or \|\or **\or \dagger\dagger
\or \ddagger\ddagger \else\@ctrerr\fi}}
\makeatother
\begin{document}

\begin{CJK*}{UTF8}{gbsn}

\title{Origin and Enhancement of Large Spin Hall Angle in Weyl Semimetals LaAl$X$ ($X$=Si, Ge)}




\author{Truman \surname{Ng} }
 \affiliation{Engineering Science Programme, National University of Singapore, Singapore 117575}

\author{Yongzheng \surname{Luo} }
\email{mpeluoy@nus.edu.sg}
\affiliation{Department of Mechanical Engineering, National University of Singapore, Singapore 117575}

\author{Jiaren \surname{Yuan} }
\email{1000005084@ujs.edu.cn}
\affiliation{Faculty of Science and School of Material Science and Engineering, Jiangsu University, Zhenjiang, China 212013}

\author{Yihong \surname{Wu} }
\affiliation{Department of Electrical and Computer Engineering, National University of Singapore, Singapore 117576}

\author{Hyunsoo \surname{Yang} }
\affiliation{Department of Electrical and Computer Engineering, National University of Singapore, Singapore 117576}

\author{Lei \surname{Shen}(沈雷)}
\email{shenlei@nus.edu.sg}
\affiliation{Engineering Science Programme, National University of Singapore, Singapore 117575}
\affiliation{Department of Mechanical Engineering, National University of Singapore, Singapore 117575}
\date{\today}

\begin{abstract}
We study the origin of the strong spin Hall effect (SHE) in a recently discovered family of Weyl semimetals, LaAl$X$ ($X$=Si, Ge) via a first-principles approach with maximally localized Wannier functions. We show that the strong intrinsic SHE in LaAl$X$ originates from the multiple slight anticrossings of nodal lines and points near $E_F$ due to their high mirror symmetry and large spin-orbit interaction. It is further found that both electrical and thermal means can enhance the spin Hall conductivity ($\sigma_{SH}$). However, the former also increases the electrical conductivity ($\sigma_{c}$), while the latter decreases it. As a result, the independent tuning of $\sigma_{SH}$ and $\sigma_{c}$ by thermal means can enhance the spin Hall angle (proportional to $\frac{\sigma_{SH}}{\sigma_{c}}$), a figure of merit of charge-to-spin current interconversion of spin-orbit torque devices. The underlying physics of such independent changes of the spin Hall and electrical conductivity by thermal means is revealed through the band-resolved and $k$-resolved spin Berry curvature. Our finding offers a new way in the search of high SHA materials for room-temperature spin-orbitronics applications.
\end{abstract}

\maketitle
\end{CJK*}

\section{\label{sec:intr} Introduction}

Spin-orbitronics is a new emerging direction of spintronics, which exploits the relativistic spin-orbit coupling (SOC), and opens fascinating new roads for spin devices made of nonmagnetic materials and operated without magnetic fields \cite{Sato2018NE,Wang2019Science,Shi2019NN,Mishra2019NC,Luo2019PRA,Liu2019,Cai2020NE,miron2011Nature,liu2012Science,manchon2019RMP}. The strong SOC allows the conversion of charge current into spin current or vice versa by the spin Hall effect (SHE) in bulk nonmagnetic materials \cite{Xu2019PRB,Zhang2017PRB,Song2020NM,MacNeill2017NP,Xu2020AM}. A direct application of charge-spin conversion is the spin-orbit torque (SOT), which utilizes the SHE to provide an ultra-fast and energy-efficient means to switch magnetization of the ferromagnets (FM) in FM/heavy-transition-metal heterostructures \cite{Liu2019,Shi2019NN,Shi2018PRB,Xu2019PRB,Wang2015PRL,MacNeill2017NP}. The SOT-based devices have been widely reported, such as magnetic random access memories and spin logic devices \cite{Mahfouzi2020PRB}. The charge-spin conversion efficiency is described by the spin Hall angle (SHA), which is written as $\theta_{SH} = \frac{\sigma_{SH}}{\sigma_{c}}$ where $\sigma_{SH}$ and $\sigma_{c}$ are the spin Hall conductivity and the electrical conductivity, respectively \cite{Zhang2017PRB,Sui2017PRB,Qiao2018PRB}. High SHA is accomplished by increasing $\sigma_{SH}$ and reducing $\sigma_{c}$ values concurrently. However, it has been difficult to control each value independently. For example, the metal Pt has very large $\sigma_{SH}$, but it is also an excellent conductor of electricity \cite{Wang2014APL}. Semimetals usually have relative low $\sigma_{c}$ because of their unique band feature near the Fermi level, but most of them do not have a giant SHC as Pt \cite{Zhou2019PRB}. It has been reported that the $\sigma_{SH}$ of semimetals can be significantly enhanced by electrical means, such as hole doping \cite{Zhou2019PRB,Sui2017PRB}. However, their electrical conductivity will be increased, too. Increasing temperature usually reduces $\sigma_{SH}$ as reported by Sun et. al. \cite{Sun2016PRL}, meanwhile, the $\sigma_{c}$ is reduced due to the strong electron scattering and electron-phonon coupling. Therefore, finding new materials with strong intrinsic SHE and independently tunable $\sigma_{SH}$ and $\sigma_{c}$ for enhancing SHA are highly desirable for the development of the state-of-the-art spin-orbit-torque technique.

Recently, topological insulators have shown high efficiency of charge-spin-current inter-conversion because of their large SOC and unique spin-momentum locking topological surface states \cite{Wang2015PRL,Fan2014NM,Shiomi2014PRLa}. However, the insulating bulk state is unavoidable, which strongly affects the performance of topological insulators in the application of spin-orbit torque, and their reported effectiveness in the experiment as spin Hall materials is debated. For example, the reported $\theta_{SH}$ are vary widely ranging from 0.0001 to 425 even though the experimental techniques are the same. \cite{Wang2015PRL,Fan2014NM,Shiomi2014PRLa}. Weyl semimetals (WSMs), the cousin of topological insulators, are conductors and have similar spin-momentum locking in both surface and bulk states \cite{Weng2015PRX,Soluyanov2015Nature}. WSMs  feature Dirac-like cones in their bulk near the Fermi energy ($E_F$) through nodal lines or points (Weyl points) on or near some crystalline planes of symmetry, such as mirror planes. Some nodal lines and points are not protected by crystalline symmetry, so these are gapped out by the SOC t hat is intrinsic in these materials. The sign of the spin Berry curvature (SBC) is opposite on either side of the gap, which cannot be cancelled out if the Fermi level is in or very close to the gap. As we will show below,  such gapped nodal lines/points generate large spin Hall conductivity as the intrinsic SHC is proportional to the integration of the SBC of the occupied bands below $E_F$. Recent theoretical studies on Weyl/Dirac semimetals, such as TaAs \cite{Sun2016PRL}, IrO$_2$ \cite{Sun2017PRB}, WTe$_2$ \cite{Zhou2019PRB}, $\beta$-W \cite{Sui2017PRB}, W$_3$Ta \cite{Derunova2019SA}, PtTe$_3$ \cite{Xu2020AM}, and ZrSiTe \cite{Yen2020PRB} show large spin Hall conductivities and spin Hall angles, and some have been verified in experiments \cite{Song2020NM,Zhao2020PRR,Shi2019NN}.

Very recently, a family of type-II Weyl semimetals in rare earth compounds was reported in the experiment \cite{Chang2018PRB,Destraz2020npjCM,Hodovanets2018PRB,Lyu2020arxiv,Puphal2020PRL,Xu2017SA}. These lanthanide based compounds have a chemical formula such as RAlX where (R = La, Ce, Pr) and (X = Si, Ge). Unlike La-based compounds, Ce- and Pr-based compounds are ferromagnetic in nature due to the strong electron correlations in the 4f orbital, which breaks the time-reversal symmetry ($\mathcal {T}$) with the anomalous Hall effect \cite{Destraz2020npjCM,Hodovanets2018PRB,Puphal2020PRL}. The nonmagnetic Weyl semimetals LaAlSi and LaAlGe have four mirror planes and thus possess many nodal lines together with Weyl points near the Fermi level \cite{Hodovanets2018PRB,Lyu2020arxiv,Xu2017SA}. Furthermore, experimental results indicate that they have moderate electrical conductivity \cite{Hodovanets2018PRB,Lyu2020arxiv}. Thus, it is naturally expected the existence of a high intrinsic SHC and large SHA in this new family of Weyl semimetals.

In this article, we investigate the intrinsic spin Hall conductivity, electrical conductivity and spin Hall angle of LaAlSi and LaAlGe using the first-principles calculations, Kubo formula, Boltzmann transport equation (BTE) and electron-phonon Wannier (EPW) approach. We indeed observe a large SHC in both WSMs, which originates from multiple slightly SOC-induced nodal gaps (anticrossings) near the Fermi level. It is further found that a small shift of the chemical potential below $E_F$ by 0.12 eV, in the form of hole doping, yields a higher SHC than that of $E_F$. The calculated electrical and spin-Hall conductivity cooperatively yield a spin Hall angle of 0.04 and 0.046 in LaAlSi and LaAlGe respectively, which is comparable with Pt. The rest of this paper is organized as follows. In \textcolor{blue}{Sec.\ref{sec:meth}}, the theory and computational details are provided. The results and discussion in \textcolor{blue}{Sec.\ref{sec:resu}} have five subsections. We first present the geometrical and electronic structures in \textcolor{blue}{Sec.\ref{sec:geom}} and \textcolor{blue}{Sec.\ref{sec:band}}. In the \textcolor{blue}{Sec.\ref{sec:SHC}} and \textcolor{blue}{Sec.\ref{sec:SBC}}, we report the spin Hall conductivity as well as band-resolved and $k$-resolved spin Berry curvatures. As LaAlSi and LaAlGe have very similar geometric and electronic structures, we particulary take LaAlGe as an example in the discussion. In the \textcolor{blue}{Sec.\ref{sec:SHA}}, we calculate the electrical conductivity through the Boltzmann transport equation with the EPW approach, and then evaluate the spin Hall angles. We finally summarize our work and draw conclusions in \textcolor{blue}{Sec.\ref{sec:conc}}.

\section{\label{sec:meth}Methodology}
Our first-principles calculations were performed using the \textsc{Quantum Espresso} package \cite{Giannozzi2009JPCM,Giannozzi2017JPCM}. A plane wave basis was used and the pseudopotential was from \textsc{pslibrary} \cite{DalCorso2014CMS}. We used a fully relativistic pseudopotential with the generalized gradient approximation (GGA) based on the projector wave augmented (PAW) method with a Perdew-Burke-Ernzerhof (PBE) functional. The Hubbard energy U of 4 eV was used for La in our calculations. The plane-wave and charge density cutoff energy is 75 Ry and 750 Ry, respectively. A \textit{k}-point grid of 8 $\times$ 8 $\times$ 8 was used in the self-consistent calculations. All structures were fully relaxed with the force on each atom was less than 10$^{-3}$ Ry/Bohr. Spin-orbit interaction was taken into account self-consistently to treat the relativistic effects. Once the self-consistent calculations were completed, the Bloch functions were Fourier transformed to the maximally localized Wannier functions (MLWFs) using the \textsc{Wannier90} package \cite{Marzari2012RMP}. The SHC was calculated using the \textsc{Berry} module on a dense 120 $\times$ 120 $\times$ 120 \textit{k}-mesh. Since the spin Berry curvature has rapid variations, adaptive smearing was used.

The intrinsic SHC was calculated via the Kubo formula, as shown below, in the clean limit. We expect the SHC in the clean limit to be given by the intrinsic SHC value due to the vanishing vertex corrections under the symmetry of $H(\textbf{k})=H(-\textbf{k})$ \cite{Guo2005PRL,Yao2005PRL,Qiao2018PRB,Ryoo2019PRB}.

\begin{equation}
\sigma_{xy}^z = e{\hbar}\int_{BZ}\frac{d\boldsymbol{k}}{(2\pi)^3} \sum_{n} f_{n\boldsymbol{k}} \Omega_{xy}^{n,z}(\boldsymbol{k}),
\end{equation}

where, $f_{n\boldsymbol{k}}$ is the Fermi-Dirac distribution function for the $n$th band at $\textbf{k}$, which includes the temperature dependency of SHC. The SHC tensor element $\sigma_{xy}^z$ describes the spin current $J_x$ with spin polarization along $z$ direction due to an incoming charge current from the $y$ direction. Other elements in the third-order tensor can be obtained by changing the mutually orthogonal Cartesian directions. $\Omega_{xy}^{n,z}(\boldsymbol{k})$ is the Berry curvature of the $n$th band as:

\begin{equation}
\Omega_{xy}^{n,z}(\boldsymbol{k}) = -\sum_{m\neq{n}} \frac{2\textrm{Im}[ \braket{n\boldsymbol{k}|j_{x}^z|m\boldsymbol{k}}\braket{m\boldsymbol{k}|v_{y}|n\boldsymbol{k}} ]}
{(\epsilon_{n\boldsymbol{k}} - \epsilon_{m\boldsymbol{k}})^2}
\end{equation}

where $j_x^z$ and $v_y$ is the spin current operator and the velocity operator.

In order to evaluate the electrical conductivity, the energy and $k$-point dependent carrier relaxation time is calculated by utilizing electron-phonon Wannier (EPW) packages \cite{Ponce2016CPC}. The scattering of electrons from the acoustic and optical phonon have been accounted for. The electron energy is computed on a 6$\times$6$\times$6 $k$-point mesh and the phonon dispersion is obtained using density functional perturbation (DFPT) method on an 3$\times$3$\times$3 $q$-point grid. Subsequently, the electron-phonon matrix elements and the imaginary part of self-energies $\Im(\sum_{nk})$ corresponding to the eigenvalues $E_{nk}$ are obtained for each electronic state of the $n^{th}$ band and the $k$ momentum. The relaxation time $\tau_{nk}$ associated with the electron-phonon interaction for each electronic state can be extracted by the formula:
\begin{equation}
\begin{split}
& \tau_{nk}^{-1}=\frac{2}{\hbar}Im(\sum\nolimits_{nk})=\frac{2\pi}{\hbar}\sum\nolimits_{m}\int_{BZ}\frac{dq}{\Omega_{BZ}}|g_{mn}(k,q)|^{2}\\
& \{[(n_q+f_{m,k+q})\delta(E_{m,k+q}-E_{n,k}-\hbar\omega_q)]+[(1+n_q-f_{m,k+q}\delta(E_{m,k+q}-E_{n,k}\hbar\omega_q)]\}
\end{split}
\end{equation}
where $g_{mn}(k,q)$ is the element of the electron-phonon coupling matrix, $m$ and $n$ are band index, $k$ and $q$ are wave vectors of the initial electronic states and phonon states, $E_{mk}$ and $\hbar\omega_q$ are energies of electrons and phones with $f_{mk}$ and $n_q$ being their distribution functions, respectively. Using the eigenvalue and the carrier relaxation time, the electronic conductivity ($\sigma$) are evaluated by utilizing the Boltzmann transport theory within the constant relaxation time approximation.

\section{\label{sec:resu}Results and discussion}

\subsection{\label{sec:geom}Geometrical structures}

\begin{figure*}[htb]
  \centering
  \includegraphics[width=0.75\textwidth]{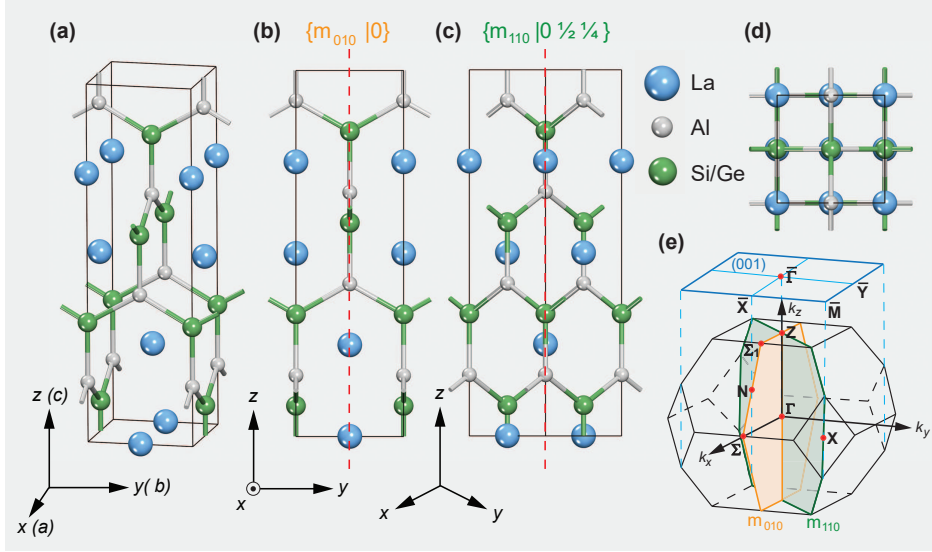}
  \caption{Crystal structures and Brilliouin zone of LaAlX (X$=$Si, Ge). (a) Conventional body-centered tetragonal structure of LaAlX with nonsymmophic space group $I$4$_1md$ (109) and $C_{4v}$ point group. $x$, $y$ and $z$ are directions along the crystal lattice of $a$, $b$ and $c$, respectively. The conventional cell possesses two pure mirror symmetries ($M_x$ and $M_y$) and two glide-mirror symmetries ($M_{xy}$ and $M_{x-y}$). Because of these symmetries, only $\{m_{010}|0\}$ and $\{m_{110}|0 \frac{1}{2} \frac{1}{4}\}$ in (b) and (c). (d) The top view of crystal. (e) The Brilliouin zone of the bulk and (001) surface with the $M_y$ and $M_{xy}$ mirror planes. The orange and turquoise green plane is invariant under mirror $M_y$ and $M_{xy}$.} %

\label{Fig.1}%
\end{figure*}

Both LaAlSi and LaAlGe have the same crystal structure, which is in a body-centered tetragonal Bravais lattice with nonsymmorphic space group $I$4$_1md$ (no. 109), $C_{4v}$ point group, lacking inversion symmetry ($\mathcal {I}$) as shown in \textcolor{blue}{\textbf{Figure 1}}. The conventional cell consists of 12 atomic layers along the (001) direction, each of which contains only one type of elements (\textcolor{blue}{\textbf{Fig. 1a}}). The $x$, $y$ and $z$ are directions along the crystal lattice of $a$, $b$ and $c$, respectively. The lattice constants are $a = b = 4.325$\AA \hspace{0.0125cm} and $c = 17.745 $\AA \hspace{0.0125cm} for LaAlSi while $a = b = 4.371 $\AA \hspace{0.0125cm} and $c = 14.849 $\AA \hspace{0.0125cm} for LaAlGe. The conventional cell possesses two pure mirror symmetries ($M_x$ and $M_y$) and two glide-mirror symmetries ($M_{xy}$ and $M_{x-y}$). Because of these symmetries, only $\{m_{010}|0\}$ and $\{m_{110}|0 \frac{1}{2} \frac{1}{4}\}$ are associated with a translation by $b/2$ and $c/4$ are shown in \textcolor{blue}{\textbf{Figs. 1b}}, \textcolor{blue}{\textbf{1c}}. The $C_2$ rotational axis can be found in the top view of \textcolor{blue}{\textbf{Fig. 1d}}. The Brillouin zone (BZ) of the bulk and (001) surface with the $M_y$ and $M_{xy}$ mirror planes (\textbf{Fig. 1e}). The orange plane, which is spanned by $\Gamma$, $\Sigma$, $N$, $\Sigma_1$ and $Z$ points, is invariant under the mirror reflection $M_y$. The turquoise green plane, which is spanned by $\Gamma$, $X$ and $Z$ points, is invariant under the mirror reflection $M_{xy}$.

\subsection{\label{sec:band}Electronic band structures}

\begin{figure*}[htbp]
\centering
\includegraphics[width=0.85\textwidth]{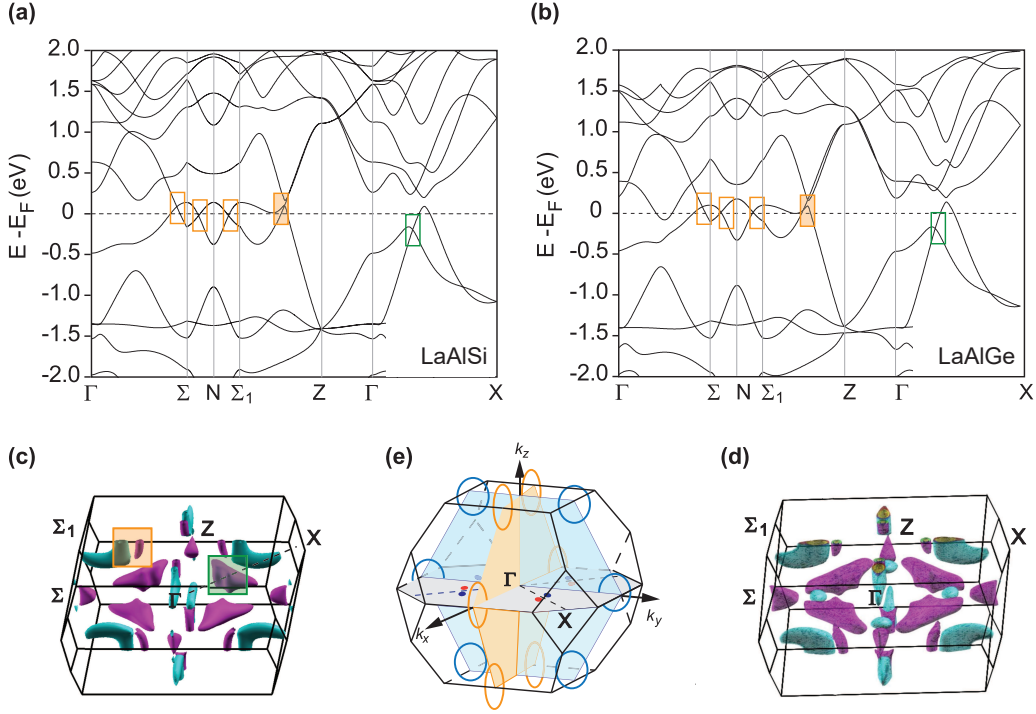} %
\caption{Band structures and the Fermi surfaces of LaAlSi and LaAlGe. (a) and (b) are band structures of LaAlSi and LaAlGe along the high-symmetry directions without SOC. Multiple Dirac-like crossings are visible near the Fermi level along the $\Gamma$-$\Sigma$-$N$-$\Sigma_1$-$Z$ lines within the $M_y$ mirror plane, and along the $\Gamma$-$X$-$Z$ lines in the $M_{xy}$ mirror plane. The crossing in the shaded orange box is a type-II Weyl node. (c) and (d) are Fermi surfaces of LaAlSi and LaAlGe over the bulk Brillouin zone. The electron- and hole-like pockets are shown in magenta and cyan colours. These band crossings form two pairs of ``nodal rings" in the two mirror planes, $M_x$ and $M_y$ in (e). Furthermore, four pairs of ``nodal points" are labelled by red and blue dots in the $k_z=0$ plane along $\Gamma$-$X$ and in the vicinity of the glide-mirror planes.} %
\label{Fig.2} %
\end{figure*}

Our band structure calculations of LaAlSi and LaAlGe without SOC are shown in \textcolor{blue}{\textbf{Figures 2a}, \textbf{2b}}. It can be seen that the conduction and valence bands cross each other along the $\Gamma-\Sigma-\Sigma_1$ path. Such Dirac-like crossings demonstrate that LaAlSi and LaAlGe are semimetals, which is in good agreement with previous reports \cite{Chang2018PRB,Hodovanets2018PRB,Xu2017SA}. It is interesting to notice the existence of two types of Weyl points, i.e., type-I and type-II. The latter is highlighted in the shaded orange box, which can be further confirmed by the calculated Fermi surface over the first bulk Brillouin zone in \textcolor{blue}{\textbf{Figs. 2c}, \textbf{2d}}. The electron- and hole-like pockets are shown in magenta and cyan colours from the Weyl cones. The e-h touching in the Fermi surface (in the shaded orange box) confirms the type-II nature of crossing in the band structure. Furthermore, we highlight a crossing along the $\Gamma$-$X$ path using a green box, the anticrossing of this point contributes to the maximum spin Hall conductivity, which will be discussed in details in the next Section. The Dirac-like crossings near to $E_F$ are mainly along the $\Gamma$-$\Sigma$-$N$-$\Sigma_{1}$-$Z$ lines and the $\Gamma$-$X$-$Z$ lines. The plane spanned by the $\Gamma$ , $N$, and $Z$ points is invariant under the mirror reflection $M_y$, and the energy bands within this plane can be labelled by the mirror eigenvalues $\pm 1$ \cite{Weng2015PRX}. The symmetry analysis in previous report\cite{Chang2018PRB} shows that the two bands that cross along the  $\Gamma$-$\Sigma$-$N$-$\Sigma_{1}$-$Z$ path belong to opposite mirror eigenvalues, and hence, the crossings between them (labelled by orange rectangles) are protected by the mirror symmetry. A similar band crossing can be found along $\Gamma$-$X$ line in the $\Gamma XZ$ plane. Together, these band crossing points form two pairs of ``nodal rings" in the two mirror planes and four pairs of ``nodal points" (\textcolor{blue}{\textbf{Fig. 2e}}). In this paper, we specially focus on the spin Hall effect in LaAlSi and LaAlGe. The detailed discussion on the Weyl features of these two materials can be found in previous computational and experimental works \cite{Chang2018PRB,Destraz2020npjCM,Hodovanets2018PRB,Lyu2020arxiv,Puphal2020PRL,Xu2017SA}.

\subsection{\label{sec:SHC}Spin Hall conductivity}

\begin{figure*}[htbp!]
  \centering
  \includegraphics[width=0.8\textwidth]{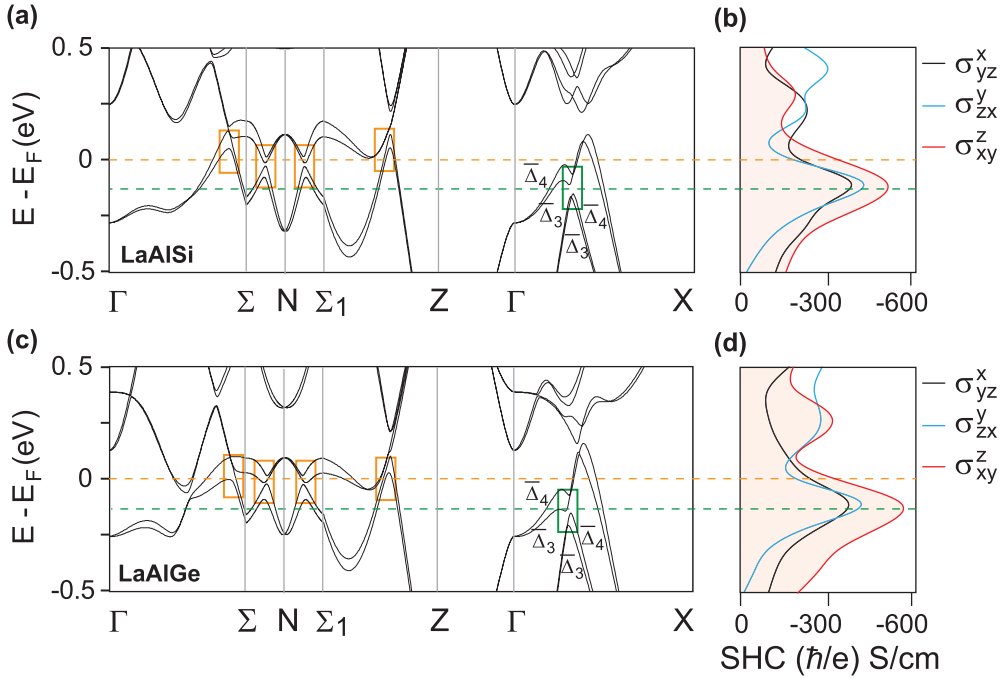}
  \newline\caption{\label{Fig.3} Relativistic band structures and spin Hall conductivities as a function of the energy of LaAlSi and LaAlGe. The Fermi level is set to zero, indicated by the orange dashed line. The green dashed line shows the position of 0.12 eV below the Fermi level. (a) and (c) The band structures of LaAlSi and LaAlGe in the presence of SOC which opens full gaps in the mirror planes because of the identical irreps ($\overline{\Delta}_{3}$ and $\overline{\Delta}_{4}$) of those Dirac-like cones. These anticrossings contribute significantly to the spin Hall conductivity. (b) and (d) The energy dependency of SHC of 3 nonzero tensor elements. The maximum SHC is marked by the green dashed line. }
\end{figure*}

In \textcolor{blue}{Sections \ref{sec:geom}, \ref{sec:band}}, we have discussed that LaAlSi and LaAlGe have 4 mirror planes and 4 other symmetry operations, such high symmetries generate multiple Dirac-like cones near $E_F$ along particular directions (forming nodal lines) or at specific points (forming nodal points). Unlike Dirac surface states in topological insulators, some robust Dirac-like points in Weyl semimetals are protected by one or more crystalline symmetries, while some are unprotected. These protected Dirac-like cones are robust against the SOC from being gapped out, such as the crossing that is described by $C_{2v}$ point group with four irreducible representations. However, those unprotected crossings would be gapped by the SOC because their ``accidental" crossings belong to the point groups with only one irreducible representation. Compared to searching for topological materials with topologically protected gapless Dirac-like cones, one should focus on  the presence of capped crossings for discovering spin Hall materials. It is because the sign of the spin berry curvature (SBC) is opposite on either side of the gap, which cannot be cancelled out if the Fermi level is in or very close to the gap. Such gaped nodal lines/points generate large spin Hall conductivity as the intrinsic SHC is proportional to the integration of the SBC of occupied bands.

In this Section, we first check how the Dirac-like crossings in the band structure (\textcolor{blue}{\textbf{Figures 2a}} and \textcolor{blue}{\textbf{2b}} ) undergo in the presence of SOC, and then calculate SHC of LaAlSi and LaAlGe. \textcolor{blue}{\textbf{Figure 3a}}  shows the band structure of LaAlSi with SOC. As shown, most Dirac-like cones, highlighted by rectangular boxes, are gapped out by SOC, which indicates that the nodal lines have disappeared in the absence of SOC. Taking the gap along $\Gamma$-$X$ in the green box as an example, we analyse the groups of bands to understand the origin of such gapping by SOC. As can be seen in \textcolor{blue}{\textbf{Fig. 3a}}, the original crossing of two bands with identical irreps $\overline{\Delta}_{3}$ and $\overline{\Delta}_{4}$, respectively, of little group $C_S$, which means SOC will gap this non-orthogonal nodal point and give rise to a non-vanishing energy term when two bands hybridize. These opened gaps near the Fermi level contribute significantly to the spin Hall conductivity. LaAlGe has the similar SOC-induced gapping features in the band structure (\textcolor{blue}{\textbf{Fig. 3c}}), and we thus expect that both of them may have large SHC according to their anticrossing features near $E_F$.

\begin{figure*}[htbp!]
  \centering
  \includegraphics[width=0.65\textwidth]{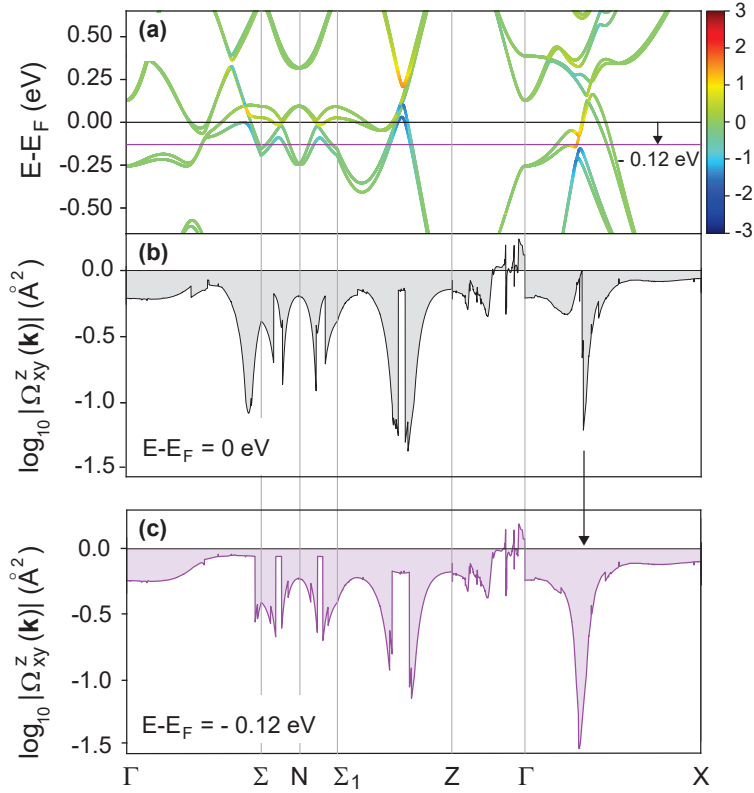}
  \newline\caption{\label{Fig.4} The band structure projected by spin Berry curvature of LaAlGe and the $k-$resolved spin Berry curvatures of $\Omega_{xy}^z$ on a log scale. (a) The SBC-projected band structure, where the red (blue) colour denotes a positive (negative) contribution of the spin Berry curvature. The solid black line and dotted purple line labels the chemical potential $\Delta \mu = 0$ eV and  $\Delta \mu = -0.12$ eV, respectively. (b) and (c) $k$-resolved SBC along the same high-symmetry paths, at $E=E_F$ and $E=E_F-0.12$ eV. A significant enhancement of the SBC peak in the $\Gamma$-$X$ segment is indicated by an arrow.}
\end{figure*}

It is worth to note that the SHC of LaAlX is anisotropic on the basis of the linear response due to their tetragonal lattice with $\mathcal {I}-$breaking. Other existing symmetries, such as mirror reflections and $\mathcal {T}$, force some tensor elements to be zero or equivalent. Thus, LaAlX only have three sets of nonzero elements $\sigma_{xy}^z=-\sigma_{yx}^z$, $\sigma_{zx}^y=-\sigma_{zy}^x$, and $\sigma_{yz}^x=-\sigma_{xz}^y$. \textcolor{blue}{\textbf{Figures 3b, 3d}} show the energy dependence of the SHC of 3 nonzero tensor elements of LaAlSi and LaAlGe. As shown, they indeed have large SHC when the chemical potential is zero ($\Delta \mu = $ 0 eV). This is summarized in \textcolor{blue}{\textbf{Table \ref{table:1}}} and compared with Pt \cite{Wang2014APL,Guo2008PRL} and some typical type-II Weyl semimetals \cite{MacNeill2017NP,Shi2019NN,Zhou2019PRB,Stiehl2019PRB}. The values of the SHC are robust as all 3 tensor elements are more than $-200 (\hbar/e)$ S/cm, while other type-II Weyl semimetals, such as WTe$_{2}$, only have a large SHC along a particular direction (see in \textcolor{blue}{\textbf{Table \ref{table:1}}}). Such strong dependence on the sample orientation for the SHE poses a challenge in experiments, and might be the reason of the large experimentally reported variation of spin Hall angles (0.029$-$0.5) in WTe$_{2}$ \cite{Shi2019NN,Zhao2020PRR,MacNeill2017NP}. Remarkably, the SHC reaches its maximum value when $E_F$ lies in the gap along the $\Gamma$-$X$ path ($\Delta \mu = $ 0.12 eV). Furthermore, the SHC$_{max}$ of LaAlGe (such as $\sigma_{xy}^z=-560$ $(\hbar/e)$ S/cm) is slightly higher than that of LaAlSi (-523 $(\hbar/e)$ S/cm) because LaAlGe has a small gap in the green box. All these results demonstrate that 1) the anticrossings by SOC in the bulk band structure are the sources of the large SHC; 2) the SHC originates from the SOC-driven gap, but its magnitude is inversely proportional to the size of the gap; and 3) the SHC can be further tuned by changing the chemical potential, such as the application of an external electric field or doping of holes.

\begin{table*}[htbp!]
\centering
\caption{\label{table:1}SHC tensor elements at the Fermi energy level ($\Delta \mu = $ 0 eV, $T=$ 0 K) in units of ($\hbar$/e) \quad S/cm, electrical conductivities in units of S/cm, and dimensionless spin Hall angles. The values in the brackets are at 300 K.}
\begin{ruledtabular}
\begin{tabular}{cccccccc}
 Material & $\sigma_{yz}^x$ & $\sigma_{zx}^y$ & $\sigma_{xy}^z$ & SHC (exp.)   &  electrical conductivity & $\mid \theta_{SH}\mid$ & Reference \\
 \hline
   Pt  (cal.)           &     &        & 2139  &             &                                  &               &  Ref.[46]     \\
   Pt  (cal.)           &     &        & 2050  &             &                                  &               &  this work    \\
   Pt  (exp.)         &     &        &          & 1900     & 5$\times 10^4$           &  0.068      &  Ref.[21]     \\
  MoTe$_2$(cal.)  & -18 & 286  & -176 &             &                                 &                & Ref.[22]       \\
  MoTe$_2$ (exp.) &      &         &        &29        & 1.8$\times 10^3$        & 0.032      & Ref.[47]       \\
  WTe$_2$ (cal.)  & -44 & 103  & -204 &             &                                  &                &   Ref.[22]     \\
  WTe$_2$ (cal.)  & 14 & 96    &         &             & 1.13$\times 10^3$      &                &  Ref.[31]      \\
   WTe$_2$ (cal.)  & -35 &     & -247    &             &                                &                &  this work      \\
  WTe$_2$ (exp.) &     &         &         &40         &  2.6$\times 10^3$       & 0.029      &  Ref.[14]      \\
  WTe$_2$ (exp.) &     &        &          & 20-300  &  1.4-1.7$\times 10^3$  & 0.09-0.5  &  Ref.[3]      \\
  LaAlSi               & -235 (-265) & -202 (-228) & -336 (-355) &           &  1.72$\times 10^4$     & 0.04        & this work      \\
  LaAlGe             & -257 (-273) & -214 (-239) & -351 (-370) &            & 1.25$\times 10^4$      &  0.06     &   this work    \\
\end{tabular}
\end{ruledtabular}
\end{table*}

\subsection{\label{sec:SBC}Spin Berry curvature}

\begin{figure*}[htbp!]
  \centering
  \includegraphics[width=0.8\textwidth]{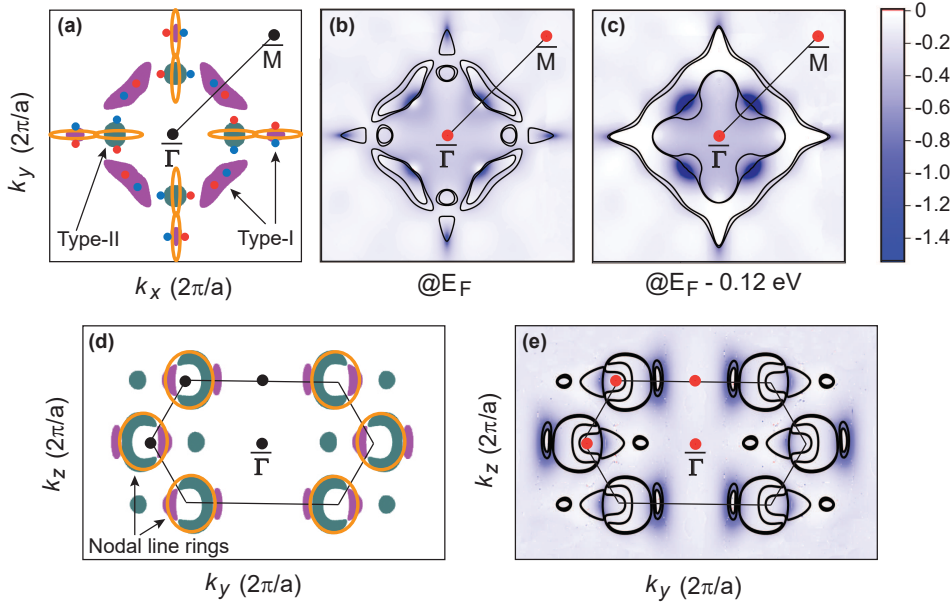}
  \newline\caption{\label{Fig.5}Projected Fermi surfaces and $k$-resolved spin Berry curvatures on a log scale in $k_x$-$k_y$ and $k_y$-$k_z$ planes in the BZ for $\sigma_{xy}^z$ SHC of LaAlGe. (a) The $k_x$-$k_y$ Fermi surface map at $k_z=0$ in Brillouin zone. The nodal rings are highlighted by orange rings. The Weyl nodal points are labelled by red/blue dots with different chirality. (b) and (c) $k$-resolved spin Berry curvatures in the 2D BZ ($k_z=0$) at $E_F$ and $E_F-0.12$ eV. (d) The $k_y$-$k_z$ Fermi surface map at $k_x=0$ in the region of BZ. The orange loops indicate a pair of nodal lines which are on the $k_y$-$k_z$ plane, protected by the mirror symmetry. (e) The $k$-resolved SBC ($k_x=0$) at $E_F$. The dominate amplitude of the spin Berry curvature (blue regions) is distributed mainly around the areas of nodal lines/points.}
\end{figure*}

In order to understand the physics of the large SHC in LaAlX and its enhancement through hole doping (shifting E$_F$ downward by 0.12 eV) in \textcolor{blue}{\textbf{Figs. 3b, 3d}}, we take $\sigma_{xy}^{z}$ of LaAlGe as an example where its spin Berry curvature is projected onto its band structure (\textcolor{blue}{\textbf{Fig. 4a}}) and plot its $k$-resolved spin Berry curvatures at $E=E_F$ and $E=E_F-0.12$ eV (\textcolor{blue}{\textbf{Figs. 4b, 4c}}). It is known that the spin Berry curvature, that is a part of a broader concept arising from the $k$-dependence of the wave function, is heavily influenced by the orbital hybridization and the position of $E_F$ in the electronic band structure. \textcolor{blue}{\textbf{Figure 4a}} shows the SBC-projected band structure, in which the red (blue) colour denotes a positive (negative) contribution of the spin Berry curvature. As shown, there are several gaps near to $E_F$ along the $\Gamma$-$N$-$Z$ lines in the Brillouin zone. Most sharp peaks in the $k$-resolved SBC (\textcolor{blue}{\textbf{Fig. 4b}}) correspond to these gaps. Thus, it is clear that the bands close to $E_F$ at the SOC-induced gapping points mainly contribute to spin Hall conductivity. This is because the unoccupied bands below the $E_F$ contribute largely to the SBC. After lowering the Fermi energy by 0.12 eV from charge neutral point ($\Delta \mu =0$ eV), the Fermi energy passes through another SOC gap along the $\Gamma$-$X$ path. From the $k$-resolved spin Berry curvature in \textcolor{blue}{\textbf{Fig. 4c}}, one can see a significant enhancement of the SBC peak between $\Gamma$ and $X$, resulting in an overall increase of SHC.

The origin of the large SBC from the gapped nodal lines/points and the increasing trend of the SBC with the shift of $E_F$ can be seen more clearly from the $k$-resolved spin Berry curvature in the 2D Brillouin zone. \textcolor{blue}{\textbf{Figures 5a and 5d }} show the 2D Fermi surface projected on the $k_x$-$k_y$ ($k_z=0$) and $k_y$-$k_z$ ($k_x=0$) plane. One can see that there are four pairs of nodal rings with two in each plane. Furthermore, a pair of nodal points are near the diagonal $\Gamma$-$M$ line. Next, let's compare the $k$-resolved Fermi surface with the $k$-resolved spin Berry curvature projected onto the same plane. As clearly shown from \textcolor{blue}{\textbf{Figs. 5a}}, \textcolor{blue}{\textbf{5b}} or \textcolor{blue}{\textbf{5d}}, \textcolor{blue}{\textbf{5e}}, the large SBC magnitude are located in the proximity of the nodal lines/points which are gapped out by SOC. Furthermore, there is a significant enhancement of the SBC around the nodal points along $\Gamma$-$M$ when $E_F$ is shifted to -0.12 eV by comparing \textcolor{blue}{\textbf{Fig. 5b}} with \textcolor{blue}{\textbf{Fig. 5c}}. The above analysis clarifies the mechanism of the SHC variation with the position of $E_F$, and sheds light on an effective approach to optimize the SHC in spin Hall materials.

\subsection{\label{sec:SHA}Spin Hall angle}

\begin{figure*}[htbp!]
  \centering
  \includegraphics[width=0.8\textwidth]{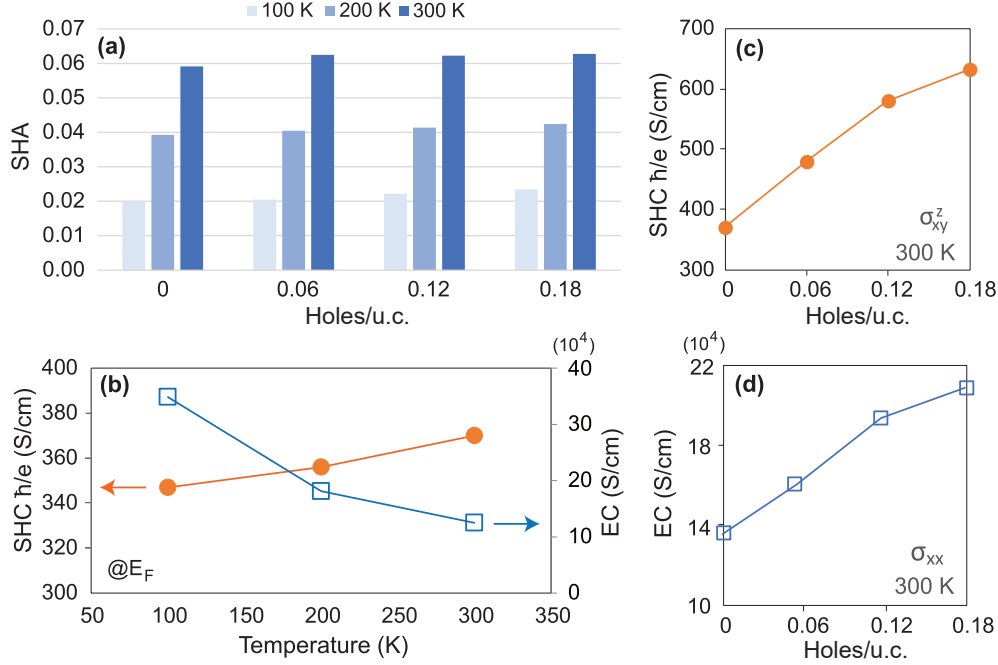}
  \newline\caption{\label{Fig.6} (a) The calculated chemical potential and temperature dependence of spin Hall angles of LaAlGe. (b) The temperature dependence of spin Hall conductivity (orange) and electrical conductivity (blue) at the Fermi lelvel (E-E$_F$=0). (c) and (d) The chemical potential dependence of spin Hall conductivity and electrical conductivity at 300 K.}
\end{figure*}

In order to determine the spin Hall angle with $\sigma_{xy}^z$, we calculate the longitudinal electrical conductivity $\sigma_{xx}$ using the Boltzmann transport equation within the constant relaxation time approximation \cite{Pizzi2014CPC}. The relaxation time was calculated using the electron-phonon Wannier method. Here, we consider two conditions, tuning the chemical potential and increasing the temperature. It is worth noting that the shifting of E$_F$ is converted into the change of the hole concentration in the calculation of electrical conductivity. We assume that a low hole doping concentration does not change the electronic band structure by much and only lowers $E_F$ slightly. The calculated hole-doping and temperature dependence of spin Hall angles of LaAlGe are shown in \textbf{Fig.6a}. As can be seen, the SHA is increases linearly with the increase in temperature. This is due to the temperature dependent changes of the SHC and electrical conductivity as shown in \textbf{Fig.6b}. The numerator $\sigma_{SH}$ (denominator $\sigma_{c}$) is increasing (decreasing) with the increase of temperature, resulting in the increase of the quotient $\theta_{SH}$ with a 300\% enhancement of $\sigma_{SH}$ from 100 K to room temperature. The increase of SHC with the increase of temperature can be understood by the calculated SHC near E$_F$ in \textbf{Fig.3d}. As shown, the Fermi level is close to the peak of SHC in LaAlGe. The smearing of the temperature will cover more states of the SHC from the peak. For comparison, the $E_F$ is located at the peak of SHC of TaAs \cite{Sun2016PRL}, the increase in temperature will cover less states, resulting in a decrease of the SHC \cite{Sun2016PRL}. In addition, it is found that the change of $E_F$ is an effective approach to enhance the SHC (\textbf{Fig.6c} and Sec.\ref{sec:SHC}). However, it will not increase the spin Hall angle due to the simultaneous increase of the electrical conductivity (\textbf{Fig.6d}). LaAlGe has a slightly larger spin Hall conductivity and smaller electrical conductivity than LaAlSi in the charge neutral case. The small charge current required in LaAlGe leads to lower Joule heating for the generation of the same amount of spin Hall current. Overall, LaAlGe has 50\% higher SHA compared with LaAlSi (see in \textcolor{blue}{\textbf{Table \ref{table:1}}}).

\section{\label{sec:conc}conclusions}
In summary, we systematically study the electronic structure, spin Hall effect, and spin Hall angle of a family of type-II Weyl semimetals, LaAlSi and LaAlGe, using the first-principles calculations with the Berry phase formalism and electron-phonon Wannier method. Both of them have large spin Hall conductivity which can be further increased by tuning the chemical potential downwards (i.e., hole doping). We reveal the physical origin of their strong SHE, which is from the high mirror symmetry and large SOC. The former gives rise to many Dirac-like crossings near the Fermi energy level, but are unprotected against SOC, resulting in anticrossings in the presence of SOC. These SOC-gapped nodal lines/points near $E_F$ create a highly unbalanced spin Berry curvature integral, and thus large spin Hall conductivity. The spin Hall conductivity can be increased by slightly tuning the Fermi level within other small gaps in the Brillouin zone. Remarkably, our results show that increasing the temperature can boost the SHC while suppressing the electrical conductivity, which enhances the spin Hall angle and the spin-orbit torque efficiency. The strong SHE can generate spin accumulation at two surfaces, which leads to unique effects in the unidirectional spin Hall magnetoresistance in FM/WSM heterostructures. The strong SHE also can be used to electrically generate spin currents and switch the magnetization of ferromagnets in spin-orbitronics applications. Furthermore, the high values of the intrinsic SHC and SHA of LaAlX could be retained and/or enhanced at room temperature, offering a great advantage for devices operated under room-temperature.

\section{\label{sec:ackn}acknowledgements}

The Authors thank Dr. Jun Zhou his for helpful discussion. K.T., H.Y. and L.S. thank MOE Tier 1 group grant (R-263-000-D60-114,R-263-000-D61-114 and R-265-000-651-114). The calculations were carried out on the GRC-NUS high-performance computing facilities.


\begin{thebibliography}{0}%
\makeatletter
\providecommand \@ifxundefined [1]{%
 \@ifx{#1\undefined}
}%
\providecommand \@ifnum [1]{%
 \ifnum #1\expandafter \@firstoftwo
 \else \expandafter \@secondoftwo
 \fi
}%
\providecommand \@ifx [1]{%
 \ifx #1\expandafter \@firstoftwo
 \else \expandafter \@secondoftwo
 \fi
}%
\providecommand \natexlab [1]{#1}%
\providecommand \enquote  [1]{``#1''}%
\providecommand \bibnamefont  [1]{#1}%
\providecommand \bibfnamefont [1]{#1}%
\providecommand \citenamefont [1]{#1}%
\providecommand \href@noop [0]{\@secondoftwo}%
\providecommand \href [0]{\begingroup \@sanitize@url \@href}%
\providecommand \@href[1]{\@@startlink{#1}\@@href}%
\providecommand \@@href[1]{\endgroup#1\@@endlink}%
\providecommand \@sanitize@url [0]{\catcode `\\12\catcode `\$12\catcode
  `\&12\catcode `\#12\catcode `\^12\catcode `\_12\catcode `\%12\relax}%
\providecommand \@@startlink[1]{}%
\providecommand \@@endlink[0]{}%
\providecommand \url  [0]{\begingroup\@sanitize@url \@url }%
\providecommand \@url [1]{\endgroup\@href {#1}{\urlprefix }}%
\providecommand \urlprefix  [0]{URL }%
\providecommand \Eprint [0]{\href }%
\providecommand \doibase [0]{https://doi.org/}%
\providecommand \selectlanguage [0]{\@gobble}%
\providecommand \bibinfo  [0]{\@secondoftwo}%
\providecommand \bibfield  [0]{\@secondoftwo}%
\providecommand \translation [1]{[#1]}%
\providecommand \BibitemOpen [0]{}%
\providecommand \bibitemStop [0]{}%
\providecommand \bibitemNoStop [0]{.\EOS\space}%
\providecommand \EOS [0]{\spacefactor3000\relax}%
\providecommand \BibitemShut  [1]{\csname bibitem#1\endcsname}%
\let\auto@bib@innerbib\@empty
\end{thebibliography}%


\begin{thebibliography}{48}%
\makeatletter
\providecommand \@ifxundefined [1]{%
 \@ifx{#1\undefined}
}%
\providecommand \@ifnum [1]{%
 \ifnum #1\expandafter \@firstoftwo
 \else \expandafter \@secondoftwo
 \fi
}%
\providecommand \@ifx [1]{%
 \ifx #1\expandafter \@firstoftwo
 \else \expandafter \@secondoftwo
 \fi
}%
\providecommand \natexlab [1]{#1}%
\providecommand \enquote  [1]{``#1''}%
\providecommand \bibnamefont  [1]{#1}%
\providecommand \bibfnamefont [1]{#1}%
\providecommand \citenamefont [1]{#1}%
\providecommand \href@noop [0]{\@secondoftwo}%
\providecommand \href [0]{\begingroup \@sanitize@url \@href}%
\providecommand \@href[1]{\@@startlink{#1}\@@href}%
\providecommand \@@href[1]{\endgroup#1\@@endlink}%
\providecommand \@sanitize@url [0]{\catcode `\\12\catcode `\$12\catcode
  `\&12\catcode `\#12\catcode `\^12\catcode `\_12\catcode `\%12\relax}%
\providecommand \@@startlink[1]{}%
\providecommand \@@endlink[0]{}%
\providecommand \url  [0]{\begingroup\@sanitize@url \@url }%
\providecommand \@url [1]{\endgroup\@href {#1}{\urlprefix }}%
\providecommand \urlprefix  [0]{URL }%
\providecommand \Eprint [0]{\href }%
\providecommand \doibase [0]{https://doi.org/}%
\providecommand \selectlanguage [0]{\@gobble}%
\providecommand \bibinfo  [0]{\@secondoftwo}%
\providecommand \bibfield  [0]{\@secondoftwo}%
\providecommand \translation [1]{[#1]}%
\providecommand \BibitemOpen [0]{}%
\providecommand \bibitemStop [0]{}%
\providecommand \bibitemNoStop [0]{.\EOS\space}%
\providecommand \EOS [0]{\spacefactor3000\relax}%
\providecommand \BibitemShut  [1]{\csname bibitem#1\endcsname}%
\let\auto@bib@innerbib\@empty
\bibitem [{\citenamefont {Sato}\ \emph {et~al.}(2018)\citenamefont {Sato},
  \citenamefont {Xue}, \citenamefont {White}, \citenamefont {Bi},\ and\
  \citenamefont {Wang}}]{Sato2018NE}%
  \BibitemOpen
  \bibfield  {author} {\bibinfo {author} {\bibfnamefont {N.}~\bibnamefont
  {Sato}}, \bibinfo {author} {\bibfnamefont {F.}~\bibnamefont {Xue}}, \bibinfo
  {author} {\bibfnamefont {R.~M.}\ \bibnamefont {White}}, \bibinfo {author}
  {\bibfnamefont {C.}~\bibnamefont {Bi}},\ and\ \bibinfo {author}
  {\bibfnamefont {S.~X.}\ \bibnamefont {Wang}},\ }\bibfield  {title} {\bibinfo
  {title} {Two-terminal spin--orbit torque magnetoresistive random access
  memory},\ }\href@noop {} {\bibfield  {journal} {\bibinfo  {journal} {Nat.
  Electron.}\ }\textbf {\bibinfo {volume} {1}},\ \bibinfo {pages} {508}
  (\bibinfo {year} {2018})}\BibitemShut {NoStop}%
\bibitem [{\citenamefont {Wang}\ \emph {et~al.}(2019)\citenamefont {Wang},
  \citenamefont {Zhu}, \citenamefont {Yang}, \citenamefont {Lee}, \citenamefont
  {Mishra}, \citenamefont {Go}, \citenamefont {Oh}, \citenamefont {Kim},
  \citenamefont {Cai}, \citenamefont {Liu} \emph {et~al.}}]{Wang2019Science}%
  \BibitemOpen
  \bibfield  {author} {\bibinfo {author} {\bibfnamefont {Y.}~\bibnamefont
  {Wang}}, \bibinfo {author} {\bibfnamefont {D.}~\bibnamefont {Zhu}}, \bibinfo
  {author} {\bibfnamefont {Y.}~\bibnamefont {Yang}}, \bibinfo {author}
  {\bibfnamefont {K.}~\bibnamefont {Lee}}, \bibinfo {author} {\bibfnamefont
  {R.}~\bibnamefont {Mishra}}, \bibinfo {author} {\bibfnamefont
  {G.}~\bibnamefont {Go}}, \bibinfo {author} {\bibfnamefont {S.-H.}\
  \bibnamefont {Oh}}, \bibinfo {author} {\bibfnamefont {D.-H.}\ \bibnamefont
  {Kim}}, \bibinfo {author} {\bibfnamefont {K.}~\bibnamefont {Cai}}, \bibinfo
  {author} {\bibfnamefont {E.}~\bibnamefont {Liu}}, \emph {et~al.},\ }\bibfield
   {title} {\bibinfo {title} {Magnetization switching by magnon-mediated spin
  torque through an antiferromagnetic insulator},\ }\href@noop {} {\bibfield
  {journal} {\bibinfo  {journal} {Science}\ }\textbf {\bibinfo {volume}
  {366}},\ \bibinfo {pages} {1125} (\bibinfo {year} {2019})}\BibitemShut
  {NoStop}%
\bibitem [{\citenamefont {Shi}\ \emph {et~al.}(2019)\citenamefont {Shi},
  \citenamefont {Liang}, \citenamefont {Zhu}, \citenamefont {Cai},
  \citenamefont {Pollard}, \citenamefont {Wang}, \citenamefont {Wang},
  \citenamefont {Wang}, \citenamefont {He}, \citenamefont {Yu} \emph
  {et~al.}}]{Shi2019NN}%
  \BibitemOpen
  \bibfield  {author} {\bibinfo {author} {\bibfnamefont {S.}~\bibnamefont
  {Shi}}, \bibinfo {author} {\bibfnamefont {S.}~\bibnamefont {Liang}}, \bibinfo
  {author} {\bibfnamefont {Z.}~\bibnamefont {Zhu}}, \bibinfo {author}
  {\bibfnamefont {K.}~\bibnamefont {Cai}}, \bibinfo {author} {\bibfnamefont
  {S.~D.}\ \bibnamefont {Pollard}}, \bibinfo {author} {\bibfnamefont
  {Y.}~\bibnamefont {Wang}}, \bibinfo {author} {\bibfnamefont {J.}~\bibnamefont
  {Wang}}, \bibinfo {author} {\bibfnamefont {Q.}~\bibnamefont {Wang}}, \bibinfo
  {author} {\bibfnamefont {P.}~\bibnamefont {He}}, \bibinfo {author}
  {\bibfnamefont {J.}~\bibnamefont {Yu}}, \emph {et~al.},\ }\bibfield  {title}
  {\bibinfo {title} {All-electric magnetization switching and
  {Dzyaloshinskii--Moriya interaction} in {WTe$_2$/ferromagnet}
  heterostructures},\ }\href@noop {} {\bibfield  {journal} {\bibinfo  {journal}
  {Nat. Nanotechnol.}\ }\textbf {\bibinfo {volume} {14}},\ \bibinfo {pages}
  {945} (\bibinfo {year} {2019})}\BibitemShut {NoStop}%
\bibitem [{\citenamefont {Mishra}\ \emph {et~al.}(2019)\citenamefont {Mishra},
  \citenamefont {Mahfouzi}, \citenamefont {Kumar}, \citenamefont {Cai},
  \citenamefont {Chen}, \citenamefont {Qiu}, \citenamefont {Kioussis},\ and\
  \citenamefont {Yang}}]{Mishra2019NC}%
  \BibitemOpen
  \bibfield  {author} {\bibinfo {author} {\bibfnamefont {R.}~\bibnamefont
  {Mishra}}, \bibinfo {author} {\bibfnamefont {F.}~\bibnamefont {Mahfouzi}},
  \bibinfo {author} {\bibfnamefont {D.}~\bibnamefont {Kumar}}, \bibinfo
  {author} {\bibfnamefont {K.}~\bibnamefont {Cai}}, \bibinfo {author}
  {\bibfnamefont {M.}~\bibnamefont {Chen}}, \bibinfo {author} {\bibfnamefont
  {X.}~\bibnamefont {Qiu}}, \bibinfo {author} {\bibfnamefont {N.}~\bibnamefont
  {Kioussis}},\ and\ \bibinfo {author} {\bibfnamefont {H.}~\bibnamefont
  {Yang}},\ }\bibfield  {title} {\bibinfo {title} {Electric-field control of
  spin accumulation direction for spin-orbit torques},\ }\href@noop {}
  {\bibfield  {journal} {\bibinfo  {journal} {Nat. Commun.}\ }\textbf {\bibinfo
  {volume} {10}},\ \bibinfo {pages} {1} (\bibinfo {year} {2019})}\BibitemShut
  {NoStop}%
\bibitem [{\citenamefont {Luo}\ \emph {et~al.}(2019)\citenamefont {Luo},
  \citenamefont {Zhang}, \citenamefont {Xu}, \citenamefont {Yang},
  \citenamefont {Zhang},\ and\ \citenamefont {Wu}}]{Luo2019PRA}%
  \BibitemOpen
  \bibfield  {author} {\bibinfo {author} {\bibfnamefont {Z.}~\bibnamefont
  {Luo}}, \bibinfo {author} {\bibfnamefont {Q.}~\bibnamefont {Zhang}}, \bibinfo
  {author} {\bibfnamefont {Y.}~\bibnamefont {Xu}}, \bibinfo {author}
  {\bibfnamefont {Y.}~\bibnamefont {Yang}}, \bibinfo {author} {\bibfnamefont
  {X.}~\bibnamefont {Zhang}},\ and\ \bibinfo {author} {\bibfnamefont
  {Y.}~\bibnamefont {Wu}},\ }\bibfield  {title} {\bibinfo {title} {Spin-orbit
  torque in a single ferromagnetic layer induced by surface spin rotation},\
  }\href {https://doi.org/10.1103/physrevapplied.11.064021} {\bibfield
  {journal} {\bibinfo  {journal} {Phys. Rev. Appl.}\ }\textbf {\bibinfo
  {volume} {11}},\ \bibinfo {pages} {064021} (\bibinfo {year}
  {2019})}\BibitemShut {NoStop}%
\bibitem [{\citenamefont {Liu}\ \emph {et~al.}(2019)\citenamefont {Liu},
  \citenamefont {Chen}, \citenamefont {Ren}, \citenamefont {Liu}, \citenamefont
  {Srivastava}, \citenamefont {Yang}, \citenamefont {Feng},\ and\ \citenamefont
  {Teo}}]{Liu2019}%
  \BibitemOpen
  \bibfield  {author} {\bibinfo {author} {\bibfnamefont {Y.}~\bibnamefont
  {Liu}}, \bibinfo {author} {\bibfnamefont {A.~P.}\ \bibnamefont {Chen}},
  \bibinfo {author} {\bibfnamefont {L.}~\bibnamefont {Ren}}, \bibinfo {author}
  {\bibfnamefont {Y.}~\bibnamefont {Liu}}, \bibinfo {author} {\bibfnamefont
  {S.}~\bibnamefont {Srivastava}}, \bibinfo {author} {\bibfnamefont
  {H.}~\bibnamefont {Yang}}, \bibinfo {author} {\bibfnamefont {Y.~P.}\
  \bibnamefont {Feng}},\ and\ \bibinfo {author} {\bibfnamefont {K.~L.}\
  \bibnamefont {Teo}},\ }\bibfield  {title} {\bibinfo {title} {Engineering
  interfacial perpendicular magnetic anisotropy in {Fe$_2$CoSi/Pt} multilayers
  with interfacial strain and orbital hybridization},\ }\href
  {https://doi.org/10.1021/acsaelm.9b00208} {\bibfield  {journal} {\bibinfo
  {journal} {ACS Appl. Electron. Mater.}\ }\textbf {\bibinfo {volume} {1}},\
  \bibinfo {pages} {1251} (\bibinfo {year} {2019})}\BibitemShut {NoStop}%
\bibitem [{\citenamefont {Cai}\ \emph {et~al.}(2020)\citenamefont {Cai},
  \citenamefont {Zhu}, \citenamefont {Lee}, \citenamefont {Mishra},
  \citenamefont {Ren}, \citenamefont {Pollard}, \citenamefont {He},
  \citenamefont {Liang}, \citenamefont {Teo},\ and\ \citenamefont
  {Yang}}]{Cai2020NE}%
  \BibitemOpen
  \bibfield  {author} {\bibinfo {author} {\bibfnamefont {K.}~\bibnamefont
  {Cai}}, \bibinfo {author} {\bibfnamefont {Z.}~\bibnamefont {Zhu}}, \bibinfo
  {author} {\bibfnamefont {J.~M.}\ \bibnamefont {Lee}}, \bibinfo {author}
  {\bibfnamefont {R.}~\bibnamefont {Mishra}}, \bibinfo {author} {\bibfnamefont
  {L.}~\bibnamefont {Ren}}, \bibinfo {author} {\bibfnamefont {S.~D.}\
  \bibnamefont {Pollard}}, \bibinfo {author} {\bibfnamefont {P.}~\bibnamefont
  {He}}, \bibinfo {author} {\bibfnamefont {G.}~\bibnamefont {Liang}}, \bibinfo
  {author} {\bibfnamefont {K.~L.}\ \bibnamefont {Teo}},\ and\ \bibinfo {author}
  {\bibfnamefont {H.}~\bibnamefont {Yang}},\ }\bibfield  {title} {\bibinfo
  {title} {Ultrafast and energy-efficient spin--orbit torque switching in
  compensated ferrimagnets},\ }\href@noop {} {\bibfield  {journal} {\bibinfo
  {journal} {Nat. Electron.}\ }\textbf {\bibinfo {volume} {3}},\ \bibinfo
  {pages} {37} (\bibinfo {year} {2020})}\BibitemShut {NoStop}%
\bibitem [{\citenamefont {Miron}\ \emph {et~al.}(2011)\citenamefont {Miron},
  \citenamefont {Garello}, \citenamefont {Gaudin}, \citenamefont {Zermatten},
  \citenamefont {Costache}, \citenamefont {Auffret}, \citenamefont {Bandiera},
  \citenamefont {Rodmacq}, \citenamefont {Schuhl},\ and\ \citenamefont
  {Gambardella}}]{miron2011Nature}%
  \BibitemOpen
  \bibfield  {author} {\bibinfo {author} {\bibfnamefont {I.~M.}\ \bibnamefont
  {Miron}}, \bibinfo {author} {\bibfnamefont {K.}~\bibnamefont {Garello}},
  \bibinfo {author} {\bibfnamefont {G.}~\bibnamefont {Gaudin}}, \bibinfo
  {author} {\bibfnamefont {P.-J.}\ \bibnamefont {Zermatten}}, \bibinfo {author}
  {\bibfnamefont {M.~V.}\ \bibnamefont {Costache}}, \bibinfo {author}
  {\bibfnamefont {S.}~\bibnamefont {Auffret}}, \bibinfo {author} {\bibfnamefont
  {S.}~\bibnamefont {Bandiera}}, \bibinfo {author} {\bibfnamefont
  {B.}~\bibnamefont {Rodmacq}}, \bibinfo {author} {\bibfnamefont
  {A.}~\bibnamefont {Schuhl}},\ and\ \bibinfo {author} {\bibfnamefont
  {P.}~\bibnamefont {Gambardella}},\ }\bibfield  {title} {\bibinfo {title}
  {Perpendicular switching of a single ferromagnetic layer induced by in-plane
  current injection},\ }\href@noop {} {\bibfield  {journal} {\bibinfo
  {journal} {Nature}\ }\textbf {\bibinfo {volume} {476}},\ \bibinfo {pages}
  {189} (\bibinfo {year} {2011})}\BibitemShut {NoStop}%
\bibitem [{\citenamefont {Liu}\ \emph {et~al.}(2012)\citenamefont {Liu},
  \citenamefont {Pai}, \citenamefont {Li}, \citenamefont {Tseng}, \citenamefont
  {Ralph},\ and\ \citenamefont {Buhrman}}]{liu2012Science}%
  \BibitemOpen
  \bibfield  {author} {\bibinfo {author} {\bibfnamefont {L.}~\bibnamefont
  {Liu}}, \bibinfo {author} {\bibfnamefont {C.-F.}\ \bibnamefont {Pai}},
  \bibinfo {author} {\bibfnamefont {Y.}~\bibnamefont {Li}}, \bibinfo {author}
  {\bibfnamefont {H.}~\bibnamefont {Tseng}}, \bibinfo {author} {\bibfnamefont
  {D.}~\bibnamefont {Ralph}},\ and\ \bibinfo {author} {\bibfnamefont
  {R.}~\bibnamefont {Buhrman}},\ }\bibfield  {title} {\bibinfo {title}
  {Spin-torque switching with the giant spin hall effect of tantalum},\
  }\href@noop {} {\bibfield  {journal} {\bibinfo  {journal} {Science}\ }\textbf
  {\bibinfo {volume} {336}},\ \bibinfo {pages} {555} (\bibinfo {year}
  {2012})}\BibitemShut {NoStop}%
\bibitem [{\citenamefont {Manchon}\ \emph {et~al.}(2019)\citenamefont
  {Manchon}, \citenamefont {{\v{Z}}elezn{\`y}}, \citenamefont {Miron},
  \citenamefont {Jungwirth}, \citenamefont {Sinova}, \citenamefont {Thiaville},
  \citenamefont {Garello},\ and\ \citenamefont {Gambardella}}]{manchon2019RMP}%
  \BibitemOpen
  \bibfield  {author} {\bibinfo {author} {\bibfnamefont {A.}~\bibnamefont
  {Manchon}}, \bibinfo {author} {\bibfnamefont {J.}~\bibnamefont
  {{\v{Z}}elezn{\`y}}}, \bibinfo {author} {\bibfnamefont {I.~M.}\ \bibnamefont
  {Miron}}, \bibinfo {author} {\bibfnamefont {T.}~\bibnamefont {Jungwirth}},
  \bibinfo {author} {\bibfnamefont {J.}~\bibnamefont {Sinova}}, \bibinfo
  {author} {\bibfnamefont {A.}~\bibnamefont {Thiaville}}, \bibinfo {author}
  {\bibfnamefont {K.}~\bibnamefont {Garello}},\ and\ \bibinfo {author}
  {\bibfnamefont {P.}~\bibnamefont {Gambardella}},\ }\bibfield  {title}
  {\bibinfo {title} {Current-induced spin-orbit torques in ferromagnetic and
  antiferromagnetic systems},\ }\href@noop {} {\bibfield  {journal} {\bibinfo
  {journal} {Reviews of Modern Physics}\ }\textbf {\bibinfo {volume} {91}},\
  \bibinfo {pages} {035004} (\bibinfo {year} {2019})}\BibitemShut {NoStop}%
\bibitem [{\citenamefont {Xu}\ \emph {et~al.}(2019)\citenamefont {Xu},
  \citenamefont {Yang}, \citenamefont {Luo},\ and\ \citenamefont
  {Wu}}]{Xu2019PRB}%
  \BibitemOpen
  \bibfield  {author} {\bibinfo {author} {\bibfnamefont {Y.}~\bibnamefont
  {Xu}}, \bibinfo {author} {\bibfnamefont {Y.}~\bibnamefont {Yang}}, \bibinfo
  {author} {\bibfnamefont {Z.}~\bibnamefont {Luo}},\ and\ \bibinfo {author}
  {\bibfnamefont {Y.}~\bibnamefont {Wu}},\ }\bibfield  {title} {\bibinfo
  {title} {Disentangling magnon magnetoresistance from anisotropic and spin
  {Hall} magnetoresistance in {NiFe/Pt} bilayers},\ }\href@noop {} {\bibfield
  {journal} {\bibinfo  {journal} {Phys. Rev. B}\ }\textbf {\bibinfo {volume}
  {100}},\ \bibinfo {pages} {094413} (\bibinfo {year} {2019})}\BibitemShut
  {NoStop}%
\bibitem [{\citenamefont {Zhang}\ \emph {et~al.}(2017)\citenamefont {Zhang},
  \citenamefont {Sun}, \citenamefont {Yang}, \citenamefont {Zelezny},
  \citenamefont {Parkin}, \citenamefont {Felser},\ and\ \citenamefont
  {Yan}}]{Zhang2017PRB}%
  \BibitemOpen
  \bibfield  {author} {\bibinfo {author} {\bibfnamefont {Y.}~\bibnamefont
  {Zhang}}, \bibinfo {author} {\bibfnamefont {Y.}~\bibnamefont {Sun}}, \bibinfo
  {author} {\bibfnamefont {H.}~\bibnamefont {Yang}}, \bibinfo {author}
  {\bibfnamefont {J.}~\bibnamefont {Zelezny}}, \bibinfo {author} {\bibfnamefont
  {S.~P.}\ \bibnamefont {Parkin}}, \bibinfo {author} {\bibfnamefont
  {C.}~\bibnamefont {Felser}},\ and\ \bibinfo {author} {\bibfnamefont
  {B.}~\bibnamefont {Yan}},\ }\bibfield  {title} {\bibinfo {title} {Strong
  anisotropic anomalous hall effect and spin hall effect in the chiral
  antiferromagnetic compounds {Mn$_3$X (X= Ge, Sn, Ga, Ir, Rh, and Pt)}},\
  }\href@noop {} {\bibfield  {journal} {\bibinfo  {journal} {Phys. Rev. B}\
  }\textbf {\bibinfo {volume} {95}},\ \bibinfo {pages} {075128} (\bibinfo
  {year} {2017})}\BibitemShut {NoStop}%
\bibitem [{\citenamefont {Song}\ \emph {et~al.}(2020)\citenamefont {Song},
  \citenamefont {Hsu}, \citenamefont {Vignale}, \citenamefont {Zhao},
  \citenamefont {Liu}, \citenamefont {Deng}, \citenamefont {Fu}, \citenamefont
  {Liu}, \citenamefont {Zhang}, \citenamefont {Lin} \emph
  {et~al.}}]{Song2020NM}%
  \BibitemOpen
  \bibfield  {author} {\bibinfo {author} {\bibfnamefont {P.}~\bibnamefont
  {Song}}, \bibinfo {author} {\bibfnamefont {C.-H.}\ \bibnamefont {Hsu}},
  \bibinfo {author} {\bibfnamefont {G.}~\bibnamefont {Vignale}}, \bibinfo
  {author} {\bibfnamefont {M.}~\bibnamefont {Zhao}}, \bibinfo {author}
  {\bibfnamefont {J.}~\bibnamefont {Liu}}, \bibinfo {author} {\bibfnamefont
  {Y.}~\bibnamefont {Deng}}, \bibinfo {author} {\bibfnamefont {W.}~\bibnamefont
  {Fu}}, \bibinfo {author} {\bibfnamefont {Y.}~\bibnamefont {Liu}}, \bibinfo
  {author} {\bibfnamefont {Y.}~\bibnamefont {Zhang}}, \bibinfo {author}
  {\bibfnamefont {H.}~\bibnamefont {Lin}}, \emph {et~al.},\ }\bibfield  {title}
  {\bibinfo {title} {Coexistence of large conventional and planar spin {Hall}
  effect with long spin diffusion length in a low-symmetry semimetal at room
  temperature},\ }\href@noop {} {\bibfield  {journal} {\bibinfo  {journal}
  {Nat. Mater.}\ }\textbf {\bibinfo {volume} {19}},\ \bibinfo {pages} {292}
  (\bibinfo {year} {2020})}\BibitemShut {NoStop}%
\bibitem [{\citenamefont {MacNeill}\ \emph {et~al.}(2017)\citenamefont
  {MacNeill}, \citenamefont {Stiehl}, \citenamefont {Guimaraes}, \citenamefont
  {Buhrman}, \citenamefont {Park},\ and\ \citenamefont
  {Ralph}}]{MacNeill2017NP}%
  \BibitemOpen
  \bibfield  {author} {\bibinfo {author} {\bibfnamefont {D.}~\bibnamefont
  {MacNeill}}, \bibinfo {author} {\bibfnamefont {G.}~\bibnamefont {Stiehl}},
  \bibinfo {author} {\bibfnamefont {M.}~\bibnamefont {Guimaraes}}, \bibinfo
  {author} {\bibfnamefont {R.}~\bibnamefont {Buhrman}}, \bibinfo {author}
  {\bibfnamefont {J.}~\bibnamefont {Park}},\ and\ \bibinfo {author}
  {\bibfnamefont {D.}~\bibnamefont {Ralph}},\ }\bibfield  {title} {\bibinfo
  {title} {Control of spin--orbit torques through crystal symmetry in
  {WTe$_2$/ferromagnet} bilayers},\ }\href@noop {} {\bibfield  {journal}
  {\bibinfo  {journal} {Nat. Phys.}\ }\textbf {\bibinfo {volume} {13}},\
  \bibinfo {pages} {300} (\bibinfo {year} {2017})}\BibitemShut {NoStop}%
\bibitem [{\citenamefont {Xu}\ \emph {et~al.}(2020)\citenamefont {Xu},
  \citenamefont {Wei}, \citenamefont {Zhou}, \citenamefont {Feng},
  \citenamefont {Xu}, \citenamefont {Du}, \citenamefont {He}, \citenamefont
  {Huang}, \citenamefont {Zhang}, \citenamefont {Liu} \emph
  {et~al.}}]{Xu2020AM}%
  \BibitemOpen
  \bibfield  {author} {\bibinfo {author} {\bibfnamefont {H.}~\bibnamefont
  {Xu}}, \bibinfo {author} {\bibfnamefont {J.}~\bibnamefont {Wei}}, \bibinfo
  {author} {\bibfnamefont {H.}~\bibnamefont {Zhou}}, \bibinfo {author}
  {\bibfnamefont {J.}~\bibnamefont {Feng}}, \bibinfo {author} {\bibfnamefont
  {T.}~\bibnamefont {Xu}}, \bibinfo {author} {\bibfnamefont {H.}~\bibnamefont
  {Du}}, \bibinfo {author} {\bibfnamefont {C.}~\bibnamefont {He}}, \bibinfo
  {author} {\bibfnamefont {Y.}~\bibnamefont {Huang}}, \bibinfo {author}
  {\bibfnamefont {J.}~\bibnamefont {Zhang}}, \bibinfo {author} {\bibfnamefont
  {Y.}~\bibnamefont {Liu}}, \emph {et~al.},\ }\bibfield  {title} {\bibinfo
  {title} {High spin {Hall} conductivity in large-area type{-II Dirac}
  semimetal {PtTe$_2$}},\ }\href@noop {} {\bibfield  {journal} {\bibinfo
  {journal} {Adv. Mater.}\ }\textbf {\bibinfo {volume} {32}},\ \bibinfo {pages}
  {2000513} (\bibinfo {year} {2020})}\BibitemShut {NoStop}%
\bibitem [{\citenamefont {Shi}\ \emph {et~al.}(2018)\citenamefont {Shi},
  \citenamefont {Wang}, \citenamefont {Wang}, \citenamefont {Ramaswamy},
  \citenamefont {Shen}, \citenamefont {Moon}, \citenamefont {Zhu},
  \citenamefont {Yu}, \citenamefont {Oh}, \citenamefont {Feng} \emph
  {et~al.}}]{Shi2018PRB}%
  \BibitemOpen
  \bibfield  {author} {\bibinfo {author} {\bibfnamefont {S.}~\bibnamefont
  {Shi}}, \bibinfo {author} {\bibfnamefont {A.}~\bibnamefont {Wang}}, \bibinfo
  {author} {\bibfnamefont {Y.}~\bibnamefont {Wang}}, \bibinfo {author}
  {\bibfnamefont {R.}~\bibnamefont {Ramaswamy}}, \bibinfo {author}
  {\bibfnamefont {L.}~\bibnamefont {Shen}}, \bibinfo {author} {\bibfnamefont
  {J.}~\bibnamefont {Moon}}, \bibinfo {author} {\bibfnamefont {D.}~\bibnamefont
  {Zhu}}, \bibinfo {author} {\bibfnamefont {J.}~\bibnamefont {Yu}}, \bibinfo
  {author} {\bibfnamefont {S.}~\bibnamefont {Oh}}, \bibinfo {author}
  {\bibfnamefont {Y.}~\bibnamefont {Feng}}, \emph {et~al.},\ }\bibfield
  {title} {\bibinfo {title} {Efficient charge-spin conversion and magnetization
  switching through the {Rashba} effect at {topological-insulator/Ag}
  interfaces},\ }\href@noop {} {\bibfield  {journal} {\bibinfo  {journal}
  {Phys. Rev. B}\ }\textbf {\bibinfo {volume} {97}},\ \bibinfo {pages} {041115}
  (\bibinfo {year} {2018})}\BibitemShut {NoStop}%
\bibitem [{\citenamefont {Wang}\ \emph {et~al.}(2015)\citenamefont {Wang},
  \citenamefont {Deorani}, \citenamefont {Banerjee}, \citenamefont {Koirala},
  \citenamefont {Brahlek}, \citenamefont {Oh},\ and\ \citenamefont
  {Yang}}]{Wang2015PRL}%
  \BibitemOpen
  \bibfield  {author} {\bibinfo {author} {\bibfnamefont {Y.}~\bibnamefont
  {Wang}}, \bibinfo {author} {\bibfnamefont {P.}~\bibnamefont {Deorani}},
  \bibinfo {author} {\bibfnamefont {K.}~\bibnamefont {Banerjee}}, \bibinfo
  {author} {\bibfnamefont {N.}~\bibnamefont {Koirala}}, \bibinfo {author}
  {\bibfnamefont {M.}~\bibnamefont {Brahlek}}, \bibinfo {author} {\bibfnamefont
  {S.}~\bibnamefont {Oh}},\ and\ \bibinfo {author} {\bibfnamefont
  {H.}~\bibnamefont {Yang}},\ }\bibfield  {title} {\bibinfo {title}
  {Topological surface states originated spin-orbit torques in
  {Bi$_2$Se$_3$}},\ }\href@noop {} {\bibfield  {journal} {\bibinfo  {journal}
  {Phys. Rev. Lett.}\ }\textbf {\bibinfo {volume} {114}},\ \bibinfo {pages}
  {257202} (\bibinfo {year} {2015})}\BibitemShut {NoStop}%
\bibitem [{\citenamefont {Mahfouzi}\ \emph {et~al.}(2020)\citenamefont
  {Mahfouzi}, \citenamefont {Mishra}, \citenamefont {Chang}, \citenamefont
  {Yang},\ and\ \citenamefont {Kioussis}}]{Mahfouzi2020PRB}%
  \BibitemOpen
  \bibfield  {author} {\bibinfo {author} {\bibfnamefont {F.}~\bibnamefont
  {Mahfouzi}}, \bibinfo {author} {\bibfnamefont {R.}~\bibnamefont {Mishra}},
  \bibinfo {author} {\bibfnamefont {P.-H.}\ \bibnamefont {Chang}}, \bibinfo
  {author} {\bibfnamefont {H.}~\bibnamefont {Yang}},\ and\ \bibinfo {author}
  {\bibfnamefont {N.}~\bibnamefont {Kioussis}},\ }\bibfield  {title} {\bibinfo
  {title} {Microscopic origin of spin-orbit torque in ferromagnetic
  heterostructures: A first-principles approach},\ }\href@noop {} {\bibfield
  {journal} {\bibinfo  {journal} {Phys. Rev. B}\ }\textbf {\bibinfo {volume}
  {101}},\ \bibinfo {pages} {060405} (\bibinfo {year} {2020})}\BibitemShut
  {NoStop}%
\bibitem [{\citenamefont {Sui}\ \emph {et~al.}(2017)\citenamefont {Sui},
  \citenamefont {Wang}, \citenamefont {Kim}, \citenamefont {Wang},
  \citenamefont {Rhim}, \citenamefont {Duan},\ and\ \citenamefont
  {Kioussis}}]{Sui2017PRB}%
  \BibitemOpen
  \bibfield  {author} {\bibinfo {author} {\bibfnamefont {X.}~\bibnamefont
  {Sui}}, \bibinfo {author} {\bibfnamefont {C.}~\bibnamefont {Wang}}, \bibinfo
  {author} {\bibfnamefont {J.}~\bibnamefont {Kim}}, \bibinfo {author}
  {\bibfnamefont {J.}~\bibnamefont {Wang}}, \bibinfo {author} {\bibfnamefont
  {S.}~\bibnamefont {Rhim}}, \bibinfo {author} {\bibfnamefont {W.}~\bibnamefont
  {Duan}},\ and\ \bibinfo {author} {\bibfnamefont {N.}~\bibnamefont
  {Kioussis}},\ }\bibfield  {title} {\bibinfo {title} {Giant enhancement of the
  intrinsic spin hall conductivity in $\beta$-tungsten via substitutional
  doping},\ }\href@noop {} {\bibfield  {journal} {\bibinfo  {journal} {Phys.
  Rev. B}\ }\textbf {\bibinfo {volume} {96}},\ \bibinfo {pages} {241105}
  (\bibinfo {year} {2017})}\BibitemShut {NoStop}%
\bibitem [{\citenamefont {Qiao}\ \emph {et~al.}(2018)\citenamefont {Qiao},
  \citenamefont {Zhou}, \citenamefont {Yuan},\ and\ \citenamefont
  {Zhao}}]{Qiao2018PRB}%
  \BibitemOpen
  \bibfield  {author} {\bibinfo {author} {\bibfnamefont {J.}~\bibnamefont
  {Qiao}}, \bibinfo {author} {\bibfnamefont {J.}~\bibnamefont {Zhou}}, \bibinfo
  {author} {\bibfnamefont {Z.}~\bibnamefont {Yuan}},\ and\ \bibinfo {author}
  {\bibfnamefont {W.}~\bibnamefont {Zhao}},\ }\bibfield  {title} {\bibinfo
  {title} {Calculation of intrinsic spin {Hall} conductivity by {Wannier}
  interpolation},\ }\href@noop {} {\bibfield  {journal} {\bibinfo  {journal}
  {Phys. Rev. B}\ }\textbf {\bibinfo {volume} {98}},\ \bibinfo {pages} {214402}
  (\bibinfo {year} {2018})}\BibitemShut {NoStop}%
\bibitem [{\citenamefont {Wang}\ \emph {et~al.}(2014)\citenamefont {Wang},
  \citenamefont {Deorani}, \citenamefont {Qiu}, \citenamefont {Kwon},\ and\
  \citenamefont {Yang}}]{Wang2014APL}%
  \BibitemOpen
  \bibfield  {author} {\bibinfo {author} {\bibfnamefont {Y.}~\bibnamefont
  {Wang}}, \bibinfo {author} {\bibfnamefont {P.}~\bibnamefont {Deorani}},
  \bibinfo {author} {\bibfnamefont {X.}~\bibnamefont {Qiu}}, \bibinfo {author}
  {\bibfnamefont {J.~H.}\ \bibnamefont {Kwon}},\ and\ \bibinfo {author}
  {\bibfnamefont {H.}~\bibnamefont {Yang}},\ }\bibfield  {title} {\bibinfo
  {title} {Determination of intrinsic spin {Hall} angle in {Pt}},\ }\href@noop
  {} {\bibfield  {journal} {\bibinfo  {journal} {Appl. Phys. Lett.}\ }\textbf
  {\bibinfo {volume} {105}},\ \bibinfo {pages} {152412} (\bibinfo {year}
  {2014})}\BibitemShut {NoStop}%
\bibitem [{\citenamefont {Zhou}\ \emph {et~al.}(2019)\citenamefont {Zhou},
  \citenamefont {Qiao}, \citenamefont {Bournel},\ and\ \citenamefont
  {Zhao}}]{Zhou2019PRB}%
  \BibitemOpen
  \bibfield  {author} {\bibinfo {author} {\bibfnamefont {J.}~\bibnamefont
  {Zhou}}, \bibinfo {author} {\bibfnamefont {J.}~\bibnamefont {Qiao}}, \bibinfo
  {author} {\bibfnamefont {A.}~\bibnamefont {Bournel}},\ and\ \bibinfo {author}
  {\bibfnamefont {W.}~\bibnamefont {Zhao}},\ }\bibfield  {title} {\bibinfo
  {title} {Intrinsic spin {Hall} conductivity of {MoTe$_2$} and {WTe$_2$}
  semimetals},\ }\href@noop {} {\bibfield  {journal} {\bibinfo  {journal}
  {Phys. Rev. B}\ }\textbf {\bibinfo {volume} {99}},\ \bibinfo {pages} {060408}
  (\bibinfo {year} {2019})}\BibitemShut {NoStop}%
\bibitem [{\citenamefont {Sun}\ \emph {et~al.}(2016)\citenamefont {Sun},
  \citenamefont {Zhang}, \citenamefont {Felser},\ and\ \citenamefont
  {Yan}}]{Sun2016PRL}%
  \BibitemOpen
  \bibfield  {author} {\bibinfo {author} {\bibfnamefont {Y.}~\bibnamefont
  {Sun}}, \bibinfo {author} {\bibfnamefont {Y.}~\bibnamefont {Zhang}}, \bibinfo
  {author} {\bibfnamefont {C.}~\bibnamefont {Felser}},\ and\ \bibinfo {author}
  {\bibfnamefont {B.}~\bibnamefont {Yan}},\ }\bibfield  {title} {\bibinfo
  {title} {Strong intrinsic spin hall effect in the {TaAs} family of {Weyl}
  semimetals},\ }\href@noop {} {\bibfield  {journal} {\bibinfo  {journal}
  {Phys. Rev. Lett.}\ }\textbf {\bibinfo {volume} {117}},\ \bibinfo {pages}
  {146403} (\bibinfo {year} {2016})}\BibitemShut {NoStop}%
\bibitem [{\citenamefont {Fan}\ \emph {et~al.}(2014)\citenamefont {Fan},
  \citenamefont {Upadhyaya}, \citenamefont {Kou}, \citenamefont {Lang},
  \citenamefont {Takei}, \citenamefont {Wang}, \citenamefont {Tang},
  \citenamefont {He}, \citenamefont {Chang}, \citenamefont {Montazeri} \emph
  {et~al.}}]{Fan2014NM}%
  \BibitemOpen
  \bibfield  {author} {\bibinfo {author} {\bibfnamefont {Y.}~\bibnamefont
  {Fan}}, \bibinfo {author} {\bibfnamefont {P.}~\bibnamefont {Upadhyaya}},
  \bibinfo {author} {\bibfnamefont {X.}~\bibnamefont {Kou}}, \bibinfo {author}
  {\bibfnamefont {M.}~\bibnamefont {Lang}}, \bibinfo {author} {\bibfnamefont
  {S.}~\bibnamefont {Takei}}, \bibinfo {author} {\bibfnamefont
  {Z.}~\bibnamefont {Wang}}, \bibinfo {author} {\bibfnamefont {J.}~\bibnamefont
  {Tang}}, \bibinfo {author} {\bibfnamefont {L.}~\bibnamefont {He}}, \bibinfo
  {author} {\bibfnamefont {L.-T.}\ \bibnamefont {Chang}}, \bibinfo {author}
  {\bibfnamefont {M.}~\bibnamefont {Montazeri}}, \emph {et~al.},\ }\bibfield
  {title} {\bibinfo {title} {Magnetization switching through giant spin--orbit
  torque in a magnetically doped topological insulator heterostructure},\
  }\href {https://doi.org/10.1038/s41535-019-0207-7} {\bibfield  {journal}
  {\bibinfo  {journal} {Nat. Mater.}\ }\textbf {\bibinfo {volume} {13}},\
  \bibinfo {pages} {699} (\bibinfo {year} {2014})}\BibitemShut {NoStop}%
\bibitem [{\citenamefont {Shiomi}\ \emph {et~al.}(2014)\citenamefont {Shiomi},
  \citenamefont {Nomura}, \citenamefont {Kajiwara}, \citenamefont {Eto},
  \citenamefont {Novak}, \citenamefont {Segawa}, \citenamefont {Ando},\ and\
  \citenamefont {Saitoh}}]{Shiomi2014PRLa}%
  \BibitemOpen
  \bibfield  {author} {\bibinfo {author} {\bibfnamefont {Y.}~\bibnamefont
  {Shiomi}}, \bibinfo {author} {\bibfnamefont {K.}~\bibnamefont {Nomura}},
  \bibinfo {author} {\bibfnamefont {Y.}~\bibnamefont {Kajiwara}}, \bibinfo
  {author} {\bibfnamefont {K.}~\bibnamefont {Eto}}, \bibinfo {author}
  {\bibfnamefont {M.}~\bibnamefont {Novak}}, \bibinfo {author} {\bibfnamefont
  {K.}~\bibnamefont {Segawa}}, \bibinfo {author} {\bibfnamefont
  {Y.}~\bibnamefont {Ando}},\ and\ \bibinfo {author} {\bibfnamefont
  {E.}~\bibnamefont {Saitoh}},\ }\bibfield  {title} {\bibinfo {title}
  {Spin-electricity conversion induced by spin injection into topological
  insulators},\ }\href@noop {} {\bibfield  {journal} {\bibinfo  {journal}
  {Phys. Rev. Lett.}\ }\textbf {\bibinfo {volume} {113}},\ \bibinfo {pages}
  {196601} (\bibinfo {year} {2014})}\BibitemShut {NoStop}%
\bibitem [{\citenamefont {Weng}\ \emph {et~al.}(2015)\citenamefont {Weng},
  \citenamefont {Fang}, \citenamefont {Fang}, \citenamefont {Bernevig},\ and\
  \citenamefont {Dai}}]{Weng2015PRX}%
  \BibitemOpen
  \bibfield  {author} {\bibinfo {author} {\bibfnamefont {H.}~\bibnamefont
  {Weng}}, \bibinfo {author} {\bibfnamefont {C.}~\bibnamefont {Fang}}, \bibinfo
  {author} {\bibfnamefont {Z.}~\bibnamefont {Fang}}, \bibinfo {author}
  {\bibfnamefont {B.~A.}\ \bibnamefont {Bernevig}},\ and\ \bibinfo {author}
  {\bibfnamefont {X.}~\bibnamefont {Dai}},\ }\bibfield  {title} {\bibinfo
  {title} {Weyl semimetal phase in noncentrosymmetric transition-metal
  monophosphides},\ }\href@noop {} {\bibfield  {journal} {\bibinfo  {journal}
  {Phys. Rev. X}\ }\textbf {\bibinfo {volume} {5}},\ \bibinfo {pages} {011029}
  (\bibinfo {year} {2015})}\BibitemShut {NoStop}%
\bibitem [{\citenamefont {Soluyanov}\ \emph {et~al.}(2015)\citenamefont
  {Soluyanov}, \citenamefont {Gresch}, \citenamefont {Wang}, \citenamefont
  {Wu}, \citenamefont {Troyer}, \citenamefont {Dai},\ and\ \citenamefont
  {Bernevig}}]{Soluyanov2015Nature}%
  \BibitemOpen
  \bibfield  {author} {\bibinfo {author} {\bibfnamefont {A.~A.}\ \bibnamefont
  {Soluyanov}}, \bibinfo {author} {\bibfnamefont {D.}~\bibnamefont {Gresch}},
  \bibinfo {author} {\bibfnamefont {Z.}~\bibnamefont {Wang}}, \bibinfo {author}
  {\bibfnamefont {Q.}~\bibnamefont {Wu}}, \bibinfo {author} {\bibfnamefont
  {M.}~\bibnamefont {Troyer}}, \bibinfo {author} {\bibfnamefont
  {X.}~\bibnamefont {Dai}},\ and\ \bibinfo {author} {\bibfnamefont {B.~A.}\
  \bibnamefont {Bernevig}},\ }\bibfield  {title} {\bibinfo {title} {{Type-II
  Weyl} semimetals},\ }\href@noop {} {\bibfield  {journal} {\bibinfo  {journal}
  {Nature}\ }\textbf {\bibinfo {volume} {527}},\ \bibinfo {pages} {495}
  (\bibinfo {year} {2015})}\BibitemShut {NoStop}%
\bibitem [{\citenamefont {Sun}\ \emph {et~al.}(2017)\citenamefont {Sun},
  \citenamefont {Zhang}, \citenamefont {Liu}, \citenamefont {Felser},
  \citenamefont {Yan}, \citenamefont {Sun}, \citenamefont {Zhang},
  \citenamefont {Liu}, \citenamefont {Felser},\ and\ \citenamefont
  {Yan}}]{Sun2017PRB}%
  \BibitemOpen
  \bibfield  {author} {\bibinfo {author} {\bibfnamefont {Y.}~\bibnamefont
  {Sun}}, \bibinfo {author} {\bibfnamefont {Y.}~\bibnamefont {Zhang}}, \bibinfo
  {author} {\bibfnamefont {C.-X.}\ \bibnamefont {Liu}}, \bibinfo {author}
  {\bibfnamefont {C.}~\bibnamefont {Felser}}, \bibinfo {author} {\bibfnamefont
  {B.}~\bibnamefont {Yan}}, \bibinfo {author} {\bibfnamefont {Y.}~\bibnamefont
  {Sun}}, \bibinfo {author} {\bibfnamefont {Y.}~\bibnamefont {Zhang}}, \bibinfo
  {author} {\bibfnamefont {C.-X.}\ \bibnamefont {Liu}}, \bibinfo {author}
  {\bibfnamefont {C.}~\bibnamefont {Felser}},\ and\ \bibinfo {author}
  {\bibfnamefont {B.}~\bibnamefont {Yan}},\ }\bibfield  {title} {\bibinfo
  {title} {Dirac nodal lines and induced spin {Hall} effect in metallic rutile
  oxides},\ }\href@noop {} {\bibfield  {journal} {\bibinfo  {journal} {Phys.
  Rev. B}\ }\textbf {\bibinfo {volume} {95}},\ \bibinfo {pages} {235104}
  (\bibinfo {year} {2017})}\BibitemShut {NoStop}%
\bibitem [{\citenamefont {Derunova}\ \emph {et~al.}(2019)\citenamefont
  {Derunova}, \citenamefont {Sun}, \citenamefont {Felser}, \citenamefont
  {Parkin}, \citenamefont {Yan},\ and\ \citenamefont {Ali}}]{Derunova2019SA}%
  \BibitemOpen
  \bibfield  {author} {\bibinfo {author} {\bibfnamefont {E.}~\bibnamefont
  {Derunova}}, \bibinfo {author} {\bibfnamefont {Y.}~\bibnamefont {Sun}},
  \bibinfo {author} {\bibfnamefont {C.}~\bibnamefont {Felser}}, \bibinfo
  {author} {\bibfnamefont {S.~S.~P.}\ \bibnamefont {Parkin}}, \bibinfo {author}
  {\bibfnamefont {B.}~\bibnamefont {Yan}},\ and\ \bibinfo {author}
  {\bibfnamefont {M.~N.}\ \bibnamefont {Ali}},\ }\bibfield  {title} {\bibinfo
  {title} {Giant intrinsic spin hall effect in {W$_3$Ta} and other {A15}
  superconductors},\ }\href {https://doi.org/10.1126/sciadv.aav8575} {\bibfield
   {journal} {\bibinfo  {journal} {Comput. Mater. Sci.}\ }\textbf {\bibinfo
  {volume} {5}},\ \bibinfo {pages} {eaav8575} (\bibinfo {year}
  {2019})}\BibitemShut {NoStop}%
\bibitem [{\citenamefont {Yen}\ and\ \citenamefont {Guo}(2020)}]{Yen2020PRB}%
  \BibitemOpen
  \bibfield  {author} {\bibinfo {author} {\bibfnamefont {Y.}~\bibnamefont
  {Yen}}\ and\ \bibinfo {author} {\bibfnamefont {G.-Y.}\ \bibnamefont {Guo}},\
  }\bibfield  {title} {\bibinfo {title} {Tunable large spin {Hall} and spin
  {Nernst} effects in the {Dirac} semimetals {ZrXY (X= Si, Ge; Y= S, Se,
  Te)}},\ }\href@noop {} {\bibfield  {journal} {\bibinfo  {journal} {Phys. Rev.
  B}\ }\textbf {\bibinfo {volume} {101}},\ \bibinfo {pages} {064430} (\bibinfo
  {year} {2020})}\BibitemShut {NoStop}%
\bibitem [{\citenamefont {Zhao}\ \emph {et~al.}(2020)\citenamefont {Zhao},
  \citenamefont {Khokhriakov}, \citenamefont {Zhang}, \citenamefont {Fu},
  \citenamefont {Karpiak}, \citenamefont {Hoque}, \citenamefont {Xu},
  \citenamefont {Jiang}, \citenamefont {Yan},\ and\ \citenamefont
  {Dash}}]{Zhao2020PRR}%
  \BibitemOpen
  \bibfield  {author} {\bibinfo {author} {\bibfnamefont {B.}~\bibnamefont
  {Zhao}}, \bibinfo {author} {\bibfnamefont {D.}~\bibnamefont {Khokhriakov}},
  \bibinfo {author} {\bibfnamefont {Y.}~\bibnamefont {Zhang}}, \bibinfo
  {author} {\bibfnamefont {H.}~\bibnamefont {Fu}}, \bibinfo {author}
  {\bibfnamefont {B.}~\bibnamefont {Karpiak}}, \bibinfo {author} {\bibfnamefont
  {A.~M.}\ \bibnamefont {Hoque}}, \bibinfo {author} {\bibfnamefont
  {X.}~\bibnamefont {Xu}}, \bibinfo {author} {\bibfnamefont {Y.}~\bibnamefont
  {Jiang}}, \bibinfo {author} {\bibfnamefont {B.}~\bibnamefont {Yan}},\ and\
  \bibinfo {author} {\bibfnamefont {S.~P.}\ \bibnamefont {Dash}},\ }\bibfield
  {title} {\bibinfo {title} {Observation of spin hall effect in {Weyl}
  semimetal {WTe$_2$} at room temperature},\ }\href@noop {} {\bibfield
  {journal} {\bibinfo  {journal} {Phys. Rev. Research}\ }\textbf {\bibinfo
  {volume} {2}},\ \bibinfo {pages} {013286} (\bibinfo {year}
  {2020})}\BibitemShut {NoStop}%
\bibitem [{\citenamefont {Chang}\ \emph {et~al.}(2018)\citenamefont {Chang},
  \citenamefont {Singh}, \citenamefont {Xu}, \citenamefont {Bian},
  \citenamefont {Huang}, \citenamefont {Hsu}, \citenamefont {Belopolski},
  \citenamefont {Alidoust}, \citenamefont {Sanchez}, \citenamefont {Zheng}
  \emph {et~al.}}]{Chang2018PRB}%
  \BibitemOpen
  \bibfield  {author} {\bibinfo {author} {\bibfnamefont {G.}~\bibnamefont
  {Chang}}, \bibinfo {author} {\bibfnamefont {B.}~\bibnamefont {Singh}},
  \bibinfo {author} {\bibfnamefont {S.-Y.}\ \bibnamefont {Xu}}, \bibinfo
  {author} {\bibfnamefont {G.}~\bibnamefont {Bian}}, \bibinfo {author}
  {\bibfnamefont {S.-M.}\ \bibnamefont {Huang}}, \bibinfo {author}
  {\bibfnamefont {C.-H.}\ \bibnamefont {Hsu}}, \bibinfo {author} {\bibfnamefont
  {I.}~\bibnamefont {Belopolski}}, \bibinfo {author} {\bibfnamefont
  {N.}~\bibnamefont {Alidoust}}, \bibinfo {author} {\bibfnamefont {D.~S.}\
  \bibnamefont {Sanchez}}, \bibinfo {author} {\bibfnamefont {H.}~\bibnamefont
  {Zheng}}, \emph {et~al.},\ }\bibfield  {title} {\bibinfo {title} {Magnetic
  and noncentrosymmetric {Weyl} fermion semimetals in the {RAlGe} family of
  compounds ({R = rare earth})},\ }\href@noop {} {\bibfield  {journal}
  {\bibinfo  {journal} {Phys. Rev. B}\ }\textbf {\bibinfo {volume} {97}},\
  \bibinfo {pages} {041104} (\bibinfo {year} {2018})}\BibitemShut {NoStop}%
\bibitem [{\citenamefont {Destraz}\ \emph {et~al.}(2020)\citenamefont
  {Destraz}, \citenamefont {Das}, \citenamefont {Tsirkin}, \citenamefont {Xu},
  \citenamefont {Neupert}, \citenamefont {Chang}, \citenamefont {Schilling},
  \citenamefont {Grushin}, \citenamefont {Kohlbrecher}, \citenamefont {Keller}
  \emph {et~al.}}]{Destraz2020npjCM}%
  \BibitemOpen
  \bibfield  {author} {\bibinfo {author} {\bibfnamefont {D.}~\bibnamefont
  {Destraz}}, \bibinfo {author} {\bibfnamefont {L.}~\bibnamefont {Das}},
  \bibinfo {author} {\bibfnamefont {S.~S.}\ \bibnamefont {Tsirkin}}, \bibinfo
  {author} {\bibfnamefont {Y.}~\bibnamefont {Xu}}, \bibinfo {author}
  {\bibfnamefont {T.}~\bibnamefont {Neupert}}, \bibinfo {author} {\bibfnamefont
  {J.}~\bibnamefont {Chang}}, \bibinfo {author} {\bibfnamefont
  {A.}~\bibnamefont {Schilling}}, \bibinfo {author} {\bibfnamefont {A.~G.}\
  \bibnamefont {Grushin}}, \bibinfo {author} {\bibfnamefont {J.}~\bibnamefont
  {Kohlbrecher}}, \bibinfo {author} {\bibfnamefont {L.}~\bibnamefont {Keller}},
  \emph {et~al.},\ }\bibfield  {title} {\bibinfo {title} {Magnetism and
  anomalous transport in the {Weyl} semimetal {PrAlGe}: possible route to axial
  gauge fields},\ }\href@noop {} {\bibfield  {journal} {\bibinfo  {journal}
  {npj Quantum Mater.}\ }\textbf {\bibinfo {volume} {5}},\ \bibinfo {pages} {1}
  (\bibinfo {year} {2020})}\BibitemShut {NoStop}%
\bibitem [{\citenamefont {Hodovanets}\ \emph {et~al.}(2018)\citenamefont
  {Hodovanets}, \citenamefont {Eckberg}, \citenamefont {Zavalij}, \citenamefont
  {Kim}, \citenamefont {Lin}, \citenamefont {Zic}, \citenamefont {Campbell},
  \citenamefont {Higgins},\ and\ \citenamefont {Paglione}}]{Hodovanets2018PRB}%
  \BibitemOpen
  \bibfield  {author} {\bibinfo {author} {\bibfnamefont {H.}~\bibnamefont
  {Hodovanets}}, \bibinfo {author} {\bibfnamefont {C.}~\bibnamefont {Eckberg}},
  \bibinfo {author} {\bibfnamefont {P.}~\bibnamefont {Zavalij}}, \bibinfo
  {author} {\bibfnamefont {H.}~\bibnamefont {Kim}}, \bibinfo {author}
  {\bibfnamefont {W.-C.}\ \bibnamefont {Lin}}, \bibinfo {author} {\bibfnamefont
  {M.}~\bibnamefont {Zic}}, \bibinfo {author} {\bibfnamefont {D.}~\bibnamefont
  {Campbell}}, \bibinfo {author} {\bibfnamefont {J.}~\bibnamefont {Higgins}},\
  and\ \bibinfo {author} {\bibfnamefont {J.}~\bibnamefont {Paglione}},\
  }\bibfield  {title} {\bibinfo {title} {Single-crystal investigation of the
  proposed {type-II Weyl} semimetal {CeAlGe}},\ }\href
  {https://doi.org/10.1103/PhysRevLett.123.016801} {\bibfield  {journal}
  {\bibinfo  {journal} {Phys. Rev. B}\ }\textbf {\bibinfo {volume} {98}},\
  \bibinfo {pages} {245132} (\bibinfo {year} {2018})}\BibitemShut {NoStop}%
\bibitem [{\citenamefont {Lyu}\ \emph {et~al.}()\citenamefont {Lyu},
  \citenamefont {Xiang}, \citenamefont {Mi}, \citenamefont {Zhao},
  \citenamefont {Wang}, \citenamefont {Liu}, \citenamefont {Chen},
  \citenamefont {Ren}, \citenamefont {Li},\ and\ \citenamefont
  {Sun}}]{Lyu2020arxiv}%
  \BibitemOpen
  \bibfield  {author} {\bibinfo {author} {\bibfnamefont {M.}~\bibnamefont
  {Lyu}}, \bibinfo {author} {\bibfnamefont {J.}~\bibnamefont {Xiang}}, \bibinfo
  {author} {\bibfnamefont {Z.}~\bibnamefont {Mi}}, \bibinfo {author}
  {\bibfnamefont {H.}~\bibnamefont {Zhao}}, \bibinfo {author} {\bibfnamefont
  {Z.}~\bibnamefont {Wang}}, \bibinfo {author} {\bibfnamefont {E.}~\bibnamefont
  {Liu}}, \bibinfo {author} {\bibfnamefont {G.}~\bibnamefont {Chen}}, \bibinfo
  {author} {\bibfnamefont {Z.}~\bibnamefont {Ren}}, \bibinfo {author}
  {\bibfnamefont {G.}~\bibnamefont {Li}},\ and\ \bibinfo {author}
  {\bibfnamefont {P.}~\bibnamefont {Sun}},\ }\bibfield  {title} {\bibinfo
  {title} {Nonsaturating magnetoresistance, anomalous {Hall} effect, and
  magnetic quantum oscillations in ferromagnetic semimetal {PrAlSi}},\
  }\href@noop {} {\bibinfo  {journal} {arXiv preprint arXiv:2001.05398}\
  }\BibitemShut {NoStop}%
\bibitem [{\citenamefont {Puphal}\ \emph {et~al.}(2020)\citenamefont {Puphal},
  \citenamefont {Pomjakushin}, \citenamefont {Kanazawa}, \citenamefont
  {Ukleev}, \citenamefont {Gawryluk}, \citenamefont {Ma}, \citenamefont
  {Naamneh}, \citenamefont {Plumb}, \citenamefont {Keller}, \citenamefont
  {Cubitt} \emph {et~al.}}]{Puphal2020PRL}%
  \BibitemOpen
\bibfield  {journal} {  }\bibfield  {author} {\bibinfo {author} {\bibfnamefont
  {P.}~\bibnamefont {Puphal}}, \bibinfo {author} {\bibfnamefont
  {V.}~\bibnamefont {Pomjakushin}}, \bibinfo {author} {\bibfnamefont
  {N.}~\bibnamefont {Kanazawa}}, \bibinfo {author} {\bibfnamefont
  {V.}~\bibnamefont {Ukleev}}, \bibinfo {author} {\bibfnamefont {D.~J.}\
  \bibnamefont {Gawryluk}}, \bibinfo {author} {\bibfnamefont {J.}~\bibnamefont
  {Ma}}, \bibinfo {author} {\bibfnamefont {M.}~\bibnamefont {Naamneh}},
  \bibinfo {author} {\bibfnamefont {N.~C.}\ \bibnamefont {Plumb}}, \bibinfo
  {author} {\bibfnamefont {L.}~\bibnamefont {Keller}}, \bibinfo {author}
  {\bibfnamefont {R.}~\bibnamefont {Cubitt}}, \emph {et~al.},\ }\bibfield
  {title} {\bibinfo {title} {Topological magnetic phase in the candidate {Weyl}
  semimetal {CeAlGe}},\ }\href@noop {} {\bibfield  {journal} {\bibinfo
  {journal} {Phys. Rev. Lett.}\ }\textbf {\bibinfo {volume} {124}},\ \bibinfo
  {pages} {017202} (\bibinfo {year} {2020})}\BibitemShut {NoStop}%
\bibitem [{\citenamefont {Xu}\ \emph {et~al.}(2017)\citenamefont {Xu},
  \citenamefont {Alidoust}, \citenamefont {Chang}, \citenamefont {Lu},
  \citenamefont {Singh}, \citenamefont {Belopolski}, \citenamefont {Sanchez},
  \citenamefont {Zhang}, \citenamefont {Bian}, \citenamefont {Zheng} \emph
  {et~al.}}]{Xu2017SA}%
  \BibitemOpen
  \bibfield  {author} {\bibinfo {author} {\bibfnamefont {S.-Y.}\ \bibnamefont
  {Xu}}, \bibinfo {author} {\bibfnamefont {N.}~\bibnamefont {Alidoust}},
  \bibinfo {author} {\bibfnamefont {G.}~\bibnamefont {Chang}}, \bibinfo
  {author} {\bibfnamefont {H.}~\bibnamefont {Lu}}, \bibinfo {author}
  {\bibfnamefont {B.}~\bibnamefont {Singh}}, \bibinfo {author} {\bibfnamefont
  {I.}~\bibnamefont {Belopolski}}, \bibinfo {author} {\bibfnamefont {D.~S.}\
  \bibnamefont {Sanchez}}, \bibinfo {author} {\bibfnamefont {X.}~\bibnamefont
  {Zhang}}, \bibinfo {author} {\bibfnamefont {G.}~\bibnamefont {Bian}},
  \bibinfo {author} {\bibfnamefont {H.}~\bibnamefont {Zheng}}, \emph {et~al.},\
  }\bibfield  {title} {\bibinfo {title} {Discovery of {Lorentz-violating}
  type{-II Weyl} fermions in {LaAlGe}},\ }\href@noop {} {\bibfield  {journal}
  {\bibinfo  {journal} {Sci. Adv.}\ }\textbf {\bibinfo {volume} {3}},\ \bibinfo
  {pages} {e1603266} (\bibinfo {year} {2017})}\BibitemShut {NoStop}%
\bibitem [{\citenamefont {Giannozzi}\ \emph {et~al.}(2009)\citenamefont
  {Giannozzi}, \citenamefont {Baroni}, \citenamefont {Bonini}, \citenamefont
  {Calandra}, \citenamefont {Car}, \citenamefont {Cavazzoni}, \citenamefont
  {Ceresoli}, \citenamefont {Chiarotti}, \citenamefont {Cococcioni},
  \citenamefont {Dabo} \emph {et~al.}}]{Giannozzi2009JPCM}%
  \BibitemOpen
  \bibfield  {author} {\bibinfo {author} {\bibfnamefont {P.}~\bibnamefont
  {Giannozzi}}, \bibinfo {author} {\bibfnamefont {S.}~\bibnamefont {Baroni}},
  \bibinfo {author} {\bibfnamefont {N.}~\bibnamefont {Bonini}}, \bibinfo
  {author} {\bibfnamefont {M.}~\bibnamefont {Calandra}}, \bibinfo {author}
  {\bibfnamefont {R.}~\bibnamefont {Car}}, \bibinfo {author} {\bibfnamefont
  {C.}~\bibnamefont {Cavazzoni}}, \bibinfo {author} {\bibfnamefont
  {D.}~\bibnamefont {Ceresoli}}, \bibinfo {author} {\bibfnamefont {G.~L.}\
  \bibnamefont {Chiarotti}}, \bibinfo {author} {\bibfnamefont {M.}~\bibnamefont
  {Cococcioni}}, \bibinfo {author} {\bibfnamefont {I.}~\bibnamefont {Dabo}},
  \emph {et~al.},\ }\bibfield  {title} {\bibinfo {title} {{QUANTUM ESPRESSO}: a
  modular and open-source software project for quantum simulations of
  materials},\ }\href {https://doi.org/10.1038/s41535-019-0207-7} {\bibfield
  {journal} {\bibinfo  {journal} {J. Phys.: Condens. Matter}\ }\textbf
  {\bibinfo {volume} {21}},\ \bibinfo {pages} {395502} (\bibinfo {year}
  {2009})}\BibitemShut {NoStop}%
\bibitem [{\citenamefont {Giannozzi}\ \emph {et~al.}(2017)\citenamefont
  {Giannozzi}, \citenamefont {Andreussi}, \citenamefont {Brumme}, \citenamefont
  {Bunau}, \citenamefont {Nardelli}, \citenamefont {Calandra}, \citenamefont
  {Car}, \citenamefont {Cavazzoni}, \citenamefont {Ceresoli}, \citenamefont
  {Cococcioni} \emph {et~al.}}]{Giannozzi2017JPCM}%
  \BibitemOpen
  \bibfield  {author} {\bibinfo {author} {\bibfnamefont {P.}~\bibnamefont
  {Giannozzi}}, \bibinfo {author} {\bibfnamefont {O.}~\bibnamefont
  {Andreussi}}, \bibinfo {author} {\bibfnamefont {T.}~\bibnamefont {Brumme}},
  \bibinfo {author} {\bibfnamefont {O.}~\bibnamefont {Bunau}}, \bibinfo
  {author} {\bibfnamefont {M.~B.}\ \bibnamefont {Nardelli}}, \bibinfo {author}
  {\bibfnamefont {M.}~\bibnamefont {Calandra}}, \bibinfo {author}
  {\bibfnamefont {R.}~\bibnamefont {Car}}, \bibinfo {author} {\bibfnamefont
  {C.}~\bibnamefont {Cavazzoni}}, \bibinfo {author} {\bibfnamefont
  {D.}~\bibnamefont {Ceresoli}}, \bibinfo {author} {\bibfnamefont
  {M.}~\bibnamefont {Cococcioni}}, \emph {et~al.},\ }\bibfield  {title}
  {\bibinfo {title} {Advanced capabilities for materials modelling with
  {Quantum ESPRESSO}},\ }\href {https://doi.org/10.1038/s41535-019-0207-7}
  {\bibfield  {journal} {\bibinfo  {journal} {J. Phys.: Condens. Matter}\
  }\textbf {\bibinfo {volume} {29}},\ \bibinfo {pages} {465901} (\bibinfo
  {year} {2017})}\BibitemShut {NoStop}%
\bibitem [{\citenamefont {Dal~Corso}(2014)}]{DalCorso2014CMS}%
  \BibitemOpen
  \bibfield  {author} {\bibinfo {author} {\bibfnamefont {A.}~\bibnamefont
  {Dal~Corso}},\ }\bibfield  {title} {\bibinfo {title} {Pseudopotentials
  periodic table: From {H} to {Pu}},\ }\href
  {https://doi.org/10.1103/PhysRevB.97.041104} {\bibfield  {journal} {\bibinfo
  {journal} {Comput. Mater. Sci.}\ }\textbf {\bibinfo {volume} {95}},\ \bibinfo
  {pages} {337} (\bibinfo {year} {2014})}\BibitemShut {NoStop}%
\bibitem [{\citenamefont {Marzari}\ \emph {et~al.}(2012)\citenamefont
  {Marzari}, \citenamefont {Mostofi}, \citenamefont {Yates}, \citenamefont
  {Souza},\ and\ \citenamefont {Vanderbilt}}]{Marzari2012RMP}%
  \BibitemOpen
  \bibfield  {author} {\bibinfo {author} {\bibfnamefont {N.}~\bibnamefont
  {Marzari}}, \bibinfo {author} {\bibfnamefont {A.~A.}\ \bibnamefont
  {Mostofi}}, \bibinfo {author} {\bibfnamefont {J.~R.}\ \bibnamefont {Yates}},
  \bibinfo {author} {\bibfnamefont {I.}~\bibnamefont {Souza}},\ and\ \bibinfo
  {author} {\bibfnamefont {D.}~\bibnamefont {Vanderbilt}},\ }\bibfield  {title}
  {\bibinfo {title} {Maximally localized {Wannier} functions: Theory and
  applications},\ }\href@noop {} {\bibfield  {journal} {\bibinfo  {journal}
  {Rev. Mod. Phys.}\ }\textbf {\bibinfo {volume} {84}},\ \bibinfo {pages}
  {1419} (\bibinfo {year} {2012})}\BibitemShut {NoStop}%
\bibitem [{\citenamefont {Guo}\ \emph {et~al.}(2005)\citenamefont {Guo},
  \citenamefont {Yao},\ and\ \citenamefont {Niu}}]{Guo2005PRL}%
  \BibitemOpen
  \bibfield  {author} {\bibinfo {author} {\bibfnamefont {G.}~\bibnamefont
  {Guo}}, \bibinfo {author} {\bibfnamefont {Y.}~\bibnamefont {Yao}},\ and\
  \bibinfo {author} {\bibfnamefont {Q.}~\bibnamefont {Niu}},\ }\bibfield
  {title} {\bibinfo {title} {Ab initio calculation of the intrinsic spin {Hall}
  effect in semiconductors},\ }\href
  {https://doi.org/10.1103/physrevlett.100.096401} {\bibfield  {journal}
  {\bibinfo  {journal} {Phys. Rev. Lett.}\ }\textbf {\bibinfo {volume} {94}},\
  \bibinfo {pages} {226601} (\bibinfo {year} {2005})}\BibitemShut {NoStop}%
\bibitem [{\citenamefont {Yao}\ and\ \citenamefont {Fang}(2005)}]{Yao2005PRL}%
  \BibitemOpen
  \bibfield  {author} {\bibinfo {author} {\bibfnamefont {Y.}~\bibnamefont
  {Yao}}\ and\ \bibinfo {author} {\bibfnamefont {Z.}~\bibnamefont {Fang}},\
  }\bibfield  {title} {\bibinfo {title} {Sign changes of intrinsic spin hall
  effect in semiconductors and simple metals: First-principles calculations},\
  }\href@noop {} {\bibfield  {journal} {\bibinfo  {journal} {Phys. Rev. Lett.}\
  }\textbf {\bibinfo {volume} {95}},\ \bibinfo {pages} {156601} (\bibinfo
  {year} {2005})}\BibitemShut {NoStop}%
\bibitem [{\citenamefont {Ryoo}\ \emph {et~al.}(2019)\citenamefont {Ryoo},
  \citenamefont {Park},\ and\ \citenamefont {Souza}}]{Ryoo2019PRB}%
  \BibitemOpen
  \bibfield  {author} {\bibinfo {author} {\bibfnamefont {J.~H.}\ \bibnamefont
  {Ryoo}}, \bibinfo {author} {\bibfnamefont {C.-H.}\ \bibnamefont {Park}},\
  and\ \bibinfo {author} {\bibfnamefont {I.}~\bibnamefont {Souza}},\ }\bibfield
   {title} {\bibinfo {title} {Computation of intrinsic spin {Hall}
  conductivities from first principles using maximally localized {Wannier}
  functions},\ }\href@noop {} {\bibfield  {journal} {\bibinfo  {journal} {Phys.
  Rev. B}\ }\textbf {\bibinfo {volume} {99}},\ \bibinfo {pages} {235113}
  (\bibinfo {year} {2019})}\BibitemShut {NoStop}%
\bibitem [{\citenamefont {Ponce}\ \emph {et~al.}(2016)\citenamefont {Ponce},
  \citenamefont {Margine}, \citenamefont {Verdi},\ and\ \citenamefont
  {Giustino}}]{Ponce2016CPC}%
  \BibitemOpen
  \bibfield  {author} {\bibinfo {author} {\bibfnamefont {S.}~\bibnamefont
  {Ponce}}, \bibinfo {author} {\bibfnamefont {E.~R.}\ \bibnamefont {Margine}},
  \bibinfo {author} {\bibfnamefont {C.}~\bibnamefont {Verdi}},\ and\ \bibinfo
  {author} {\bibfnamefont {F.}~\bibnamefont {Giustino}},\ }\bibfield  {title}
  {\bibinfo {title} {Epw: Electron--phonon coupling, transport and
  superconducting properties using maximally localized {Wannier} functions},\
  }\href@noop {} {\bibfield  {journal} {\bibinfo  {journal} {Comput. Phys.
  Commun.}\ }\textbf {\bibinfo {volume} {209}},\ \bibinfo {pages} {116}
  (\bibinfo {year} {2016})}\BibitemShut {NoStop}%
\bibitem [{\citenamefont {Guo}\ \emph {et~al.}(2008)\citenamefont {Guo},
  \citenamefont {Murakami}, \citenamefont {Chen},\ and\ \citenamefont
  {Nagaosa}}]{Guo2008PRL}%
  \BibitemOpen
  \bibfield  {author} {\bibinfo {author} {\bibfnamefont {G.~Y.}\ \bibnamefont
  {Guo}}, \bibinfo {author} {\bibfnamefont {S.}~\bibnamefont {Murakami}},
  \bibinfo {author} {\bibfnamefont {T.-W.}\ \bibnamefont {Chen}},\ and\
  \bibinfo {author} {\bibfnamefont {N.}~\bibnamefont {Nagaosa}},\ }\bibfield
  {title} {\bibinfo {title} {Intrinsic spin {Hall} effect in {Platinum}:
  First-principles calculations},\ }\href
  {https://doi.org/10.1103/physrevlett.100.096401} {\bibfield  {journal}
  {\bibinfo  {journal} {Phys. Rev. Lett.}\ }\textbf {\bibinfo {volume} {100}},\
  \bibinfo {pages} {465901} (\bibinfo {year} {2008})}\BibitemShut {NoStop}%
\bibitem [{\citenamefont {Stiehl}\ \emph {et~al.}(2019)\citenamefont {Stiehl},
  \citenamefont {Li}, \citenamefont {Gupta}, \citenamefont {El~Baggari},
  \citenamefont {Jiang}, \citenamefont {Xie}, \citenamefont {Kourkoutis},
  \citenamefont {Mak}, \citenamefont {Shan}, \citenamefont {Buhrman} \emph
  {et~al.}}]{Stiehl2019PRB}%
  \BibitemOpen
  \bibfield  {author} {\bibinfo {author} {\bibfnamefont {G.~M.}\ \bibnamefont
  {Stiehl}}, \bibinfo {author} {\bibfnamefont {R.}~\bibnamefont {Li}}, \bibinfo
  {author} {\bibfnamefont {V.}~\bibnamefont {Gupta}}, \bibinfo {author}
  {\bibfnamefont {I.}~\bibnamefont {El~Baggari}}, \bibinfo {author}
  {\bibfnamefont {S.}~\bibnamefont {Jiang}}, \bibinfo {author} {\bibfnamefont
  {H.}~\bibnamefont {Xie}}, \bibinfo {author} {\bibfnamefont {L.~F.}\
  \bibnamefont {Kourkoutis}}, \bibinfo {author} {\bibfnamefont {K.~F.}\
  \bibnamefont {Mak}}, \bibinfo {author} {\bibfnamefont {J.}~\bibnamefont
  {Shan}}, \bibinfo {author} {\bibfnamefont {R.~A.}\ \bibnamefont {Buhrman}},
  \emph {et~al.},\ }\bibfield  {title} {\bibinfo {title} {Layer-dependent
  spin-orbit torques generated by the centrosymmetric transition metal
  dichalcogenide {$\beta$- MoTe$_2$}},\ }\href@noop {} {\bibfield  {journal}
  {\bibinfo  {journal} {Phys. Rev. B}\ }\textbf {\bibinfo {volume} {100}},\
  \bibinfo {pages} {184402} (\bibinfo {year} {2019})}\BibitemShut {NoStop}%
\bibitem [{\citenamefont {Pizzi}\ \emph {et~al.}(2014)\citenamefont {Pizzi},
  \citenamefont {Volja}, \citenamefont {Kozinsky}, \citenamefont {Fornari},\
  and\ \citenamefont {Marzari}}]{Pizzi2014CPC}%
  \BibitemOpen
  \bibfield  {author} {\bibinfo {author} {\bibfnamefont {G.}~\bibnamefont
  {Pizzi}}, \bibinfo {author} {\bibfnamefont {D.}~\bibnamefont {Volja}},
  \bibinfo {author} {\bibfnamefont {B.}~\bibnamefont {Kozinsky}}, \bibinfo
  {author} {\bibfnamefont {M.}~\bibnamefont {Fornari}},\ and\ \bibinfo {author}
  {\bibfnamefont {N.}~\bibnamefont {Marzari}},\ }\bibfield  {title} {\bibinfo
  {title} {Boltzwann: A code for the evaluation of thermoelectric and
  electronic transport properties with a maximally-localized {Wannier}
  functions basis},\ }\href@noop {} {\bibfield  {journal} {\bibinfo  {journal}
  {Comput. Phys. Commun.}\ }\textbf {\bibinfo {volume} {185}},\ \bibinfo
  {pages} {422} (\bibinfo {year} {2014})}\BibitemShut {NoStop}%
\end{thebibliography}
%

\end{document}